\documentclass[12pt]{article}
\pdfoutput=1
\usepackage{jheppub}

\usepackage{slashed}
\usepackage{color, verbatim}
\usepackage{latexsym}
\usepackage{epsfig}
\usepackage{amsmath,accents}

\newcommand*{\dt}[1]{%
  \accentset{\mbox{\bfseries .}}{#1}}
\newcommand*{\ddt}[1]{%
  \accentset{\mbox{\bfseries .\hspace{-0.25ex}.}}{#1}}
\usepackage{amssymb}
\usepackage{graphicx}
\usepackage{bm}
\usepackage{stackrel}

\usepackage[font={small}]{caption}

\usepackage{hyperref}
\usepackage{epstopdf}
\definecolor{myred}{rgb}{0.7, 0, 0}
\definecolor{myblue}{rgb}{0, 0, 0.7}
\definecolor{mygreen}{rgb}{0.04, 0.7, 0.5}
\hypersetup{colorlinks,citecolor=myred,linkcolor=myblue,urlcolor=myblue,linktocpage=true}

\makeatletter

\@addtoreset{equation}{section}
\makeatother

\newcommand{\be}{\begin{equation}}
\newcommand{\ee}{\end{equation}}
\newcommand{\bea}{\begin{eqnarray}}
\newcommand{\eea}{\end{eqnarray}}

\newcommand{\diag}{\operatorname{diag}}

\newcommand{\h}{\mathfrak h}

\newcommand{\TeV}{{\textrm{TeV}}}
\newcommand{\cW}{{\mathcal{W}}}

\begin{document}

\thispagestyle{empty}

%\begin{comment}
\begin{center}

\hfill UAB-FT-779

\begin{center}

\vspace{.5cm}

{\Large\bf
Gapped Continuum Kaluza-Klein spectrum
\vspace{0.3cm}
}\\

\end{center}

\vspace{1.cm}

\textbf{
Eugenio Meg\'ias$^{\,a}$, Mariano Quir\'os$^{\,b}$
}\\

\vspace{.1cm}
${}^a\!\!$ {\em {Departamento de F\'{\i}sica At\'omica, Molecular y Nuclear and \\ Instituto Carlos I de F\'{\i}sica Te\'orica y Computacional, Universidad de Granada,\\ Avenida de Fuente Nueva s/n,  18071 Granada, Spain}}

${}^b\!\!$ {\em {Institut de F\'{\i}sica d'Altes Energies (IFAE) and\\ The Barcelona Institute of  Science and Technology (BIST),\\ Campus UAB, 08193 Bellaterra (Barcelona) Spain
}}

\end{center}

\vspace{0.8cm}

\centerline{\bf Abstract}
\vspace{2 mm}

\begin{quote}\small
  We consider a warped five-dimensional model with an ultraviolet (UV) brane and, on top of the Standard Model isolated modes, continua of KK modes with different mass gaps for all particles: gauge bosons, fermions, graviton, radion and Higgs boson. The model can be considered as a modelization in five dimensions of gapped unparticles. The five dimensional metric has a singularity, at a finite (infinite) value of the proper (conformal) coordinate, which is admissible as it supports finite temperature in the form of a black hole horizon. An infrared (IR) brane, with particular jumping conditions, is introduced to trigger correct electroweak breaking. The gravitational metric is AdS$_5$ near the UV brane, to solve the hierarchy problem with a fundamental Planck scale, and linear, in conformal coordinates, near the IR, as in the linear dilaton and five-dimensional clockwork models. The branes, and singularity, distances are fixed, \textit{\` a la} Goldberger-Wise, by a bulk scalar field with brane potentials explicitly breaking the conformal symmetry. The bosonic continuum of KK modes with the smallest mass gap are those of gauge bosons, and so they are the most likely produced at the LHC. Mass gaps of the continuum of KK fermions do depend on their localization in the extra dimension. We have computed the spectral functions, and arbitrary Green's functions, and shown how they can modify some Standard Model processes. 
  
\end{quote}

\vfill

\newpage

\tableofcontents

\newpage
\section{Introduction}
\label{sec:introduction}
The Standard Model (SM) of electroweak (EW) and strong interactions has been put on solid grounds by past and current experimental data, collected at e.g.~the Large Hadron Collider (LHC) or the Large Electron Positron (LEP) collider~\cite{ALEPH:2005ab,Olive:2016xmw}. However, based on experimental information (as e.g.~the existence of dark matter), and theoretical input (as e.g.~the sensitivity of the EW scale to ultraviolet (UV) physics, a.k.a. the hierarchy problem), we believe that the SM is not the ultimate fundamental theory but, instead, it is an effective theory which works at scales below a few TeV. In order to cope with the hierarchy problem, extensions of the SM have been proposed where the above sensitivity cancels, the most popular ones being supersymmetry and theories with a warped extra dimension~\cite{Randall:1999ee}, the latter being conjectured to be dual (the AdS/CFT conjecture) to conformal four dimensional (4D) theories, with composite Higgs boson and heavy fermions, as well as towers of composite resonances. 

 In this paper we are mainly concerned for theories which solve the hierarchy problem by means of a warped extra dimension and offer, as stated above, a very interesting dual interpretation.  However the elusiveness of (isolated and narrow) heavy resonances at the LHC~\cite{Sirunyan:2018ryr,Aaboud:2019roo} has led people to imagine different solutions to the hierarchy problem that could possibly escape present detection. One possible alternative approach is the clockwork models~\cite{Giudice:2017suc}, or better their five-dimensional (5D) continuum limit~\cite{Giudice:2017fmj}, the linear dilaton models~\cite{Antoniadis:2011qw}, dual to Little String theories (LST)~\cite{Antoniadis:2001sw}, which predict an (almost) continuum spectrum with a TeV mass gap and a mass separation between modes $\sim 30$ GeV. In these models the ``fundamental" scales, i.e.~the 5D Planck scale $M$ as well as the curvature of the 5D space $k$, are at the TeV, while the 4D Planck scale, $M_P$, which is obtained after warping, is not fundamental. In these theories the weakness of gravity and the hierarchy problem are related to the smallness of the string coupling in LST. 
 
 Here we will pursue a more conventional approach and propose a model where the 4D Planck scale is the fundamental one, and the TeV gap scale is obtained by the warp factor from the UV to the IR  brane. The spectrum of particles has a TeV mass gap followed by a continuum of states. This is achieved by the bulk dynamics of a stabilizing scalar field which back reacts on the 5D gravitational metric and generates a singularity at a finite distance in proper coordinates from the UV brane. The singularity is admissible in the sense of Ref.~\cite{Gubser:2000nd}, since it satisfies the condition for a bulk geometry to support finite temperature in the form of a black hole horizon, as was proven in Ref.~\cite{Cabrer:2009we}. The bulk dynamics of the scalar field generates a 5D metric of the so-called soft-wall type. In this class of models, as the 5D metric goes to zero at the singularity, the Higgs profile has a maximum away from the IR, where the Kaluza-Klein (KK) modes are localized, which suppresses their contribution to EW observables~\cite{Falkowski:2008fz,Cabrer:2010si,Cabrer:2011fb,Cabrer:2011vu,Cabrer:2011qb,Quiros:2013yaa,Carmona:2011ib,deBlas:2012qf}, a phenomenon which was already observed in Ref.~\cite{Carmona:2011ib}, where this class of models were dubbed non-custodial models. This mechanism provides an alternative to the so-called custodial models where the bulk gauge symmetry is enlarged, to encompass custodial symmetry, to $SU(2)_L\otimes SU(2)_R\otimes U(1)_{B-L}$, which is broken to $SU(2)_L\otimes U(1)_Y$ at the UV brane but conserved at the IR brane~\cite{Agashe:2003zs}, or even custodial models where the symmetry in the bulk is an enlarged group $\mathcal G$~\cite{Agashe:2004rs}, and where the Higgs is identified with the fifth component of the 5D gauge field (dubbed gauge-Higgs unification, or composite, Higgs models) and where EW symmetry breaking proceeds dynamically.

 In this paper we will present the critical case of non-custodial models where the spectrum is continuum with a gap at the TeV scale, and the hierarchy problem is solved by a stabilizing scalar field \textit{\`a la} Goldberger-Wise~\cite{Goldberger:1999uk}. The back reaction of the scalar field on the metric generates a linear dilaton \textit{only in the IR region}, while in the UV the behavior is AdS$_5$, which allows to solve conventionally the hierarchy problem. Moreover, this permits a holographic interpretation of the model and connections with unparticles~\cite{Georgi:2007ek,Georgi:2007si} in the presence of a mass gap. The model is defined in the finite interval of the extra dimension between the UV brane and the singularity. On top of that we introduce an IR brane (the EW breaking brane) with the sole purpose of triggering EW symmetry breaking. The model has to be imposed boundary conditions on the UV brane, jump conditions on the IR brane and regularity conditions at the singularity.  

Let us stress that the idea of models with an isolated resonance and a gapped continuum spectrum, reminiscent of unparticle models~\cite{Georgi:2007ek,Georgi:2007si}, is by far not new. In fact introducing a TeV IR cutoff in a conformal theory was already proposed in Refs.~\cite{Cacciapaglia:2007jq,Cacciapaglia:2008ns}, where the gap $\mu$ was triggered by the coupling $\Phi\phi\phi$ of a field with the profile $\Phi=\mu^2 z^2$, and in principle the scale $\mu$ should be at the TeV value. Our formalism departs from this idea as the gap is induced from the fundamental scale $k$ by the warp factor, and thus is linked to the solution of the hierarchy problem. Moreover in our theory, the Higgs boson shares in particular the property of having an isolated narrow resonance, with a mass fixed by the experimental Higgs mass (triggered by the Higgs potential localized at the IR brane), and a continuum of states separated from the resonance by a TeV mass gap. In this sense our theory is a modelization of theories, dubbed Unhiggs theories, which share those features. The properties and phenomenology of Unhiggs theories were extensively studied and developed in a number of papers, as can be seen in Refs.~\cite{Stancato:2008mp,Falkowski:2008yr,Falkowski:2009uy,Stancato:2010ay,Englert:2012dq,Englert:2012cb,Bellazzini:2015cgj}. In our model we have explored the simplest possibility where the Higgs is a mesonic doublet, an additional IR brane was shown to be necessary to trigger electroweak symmetry breaking, and the stability of the IR brane was implemented by the stabilizing bulk field $\phi$ with a Goldberger-Wise mechanism explicitly breaking the conformal invariance with potentials in both branes. Of course since our IR brane is not a boundary, jump conditions guaranteeing the continuity of the solutions need to be imposed for all degrees of freedom propagating in the bulk. Moreover in our theory, not only the Higgs has its continuum excitations in the conformal sector, but also all the rest of Standard Model fields, including the gauge bosons, graviton, radion and the different fermions. Along the same direction, a linear dilaton 5D composite Higgs model with continuum spectrum has been recently analyzed in Refs.~\cite{Csaki:2018kxb,Lee:2018qte}.

The outline of the paper is as follows. We introduce in Sec.~\ref{sec:general} the general formalism for the 5D action, and the gravitational background which will be used in the rest of the paper. In particular our metric and dilaton behave linearly in conformal coordinates in the deep IR while their behavior is $AdS_5$ near the UV. The stabilization mechanism of the model, \textit{\` a la} Goldberger-Wise, is described in Sec.~\ref{sec:Brane_stabilization}. The Higgs sector of the theory, and the electroweak symmetry breaking, are then studied in Sec.~\ref{sec:Higgs}, where we also confront the model predictions with electroweak precision tests. We study in Sec.~\ref{sec:continuum_spectrum} the spectral functions and the holographic (UV-brane-to-UV-brane) Green's functions of the continuum spectra of the KK modes for all particles: gauge bosons, fermions, graviton, radion and Higgs boson. We analyze in Sec.~\ref{sec:Greens_functions} the general and, in particular, the brane-to-brane Green's functions, and study how they modify the LHC phenomenology. Finally we conclude with a discussion of our results, and an outlook toward future directions in Sec.~\ref{sec:conclusions}.

\section{The gravitational background}
\label{sec:general}

We consider a slice of 5D space-time between a brane at the value $y = y_0=0$ in proper coordinates, the UV brane, and an admissible singularity placed at $y = y_s$. In addition, we will introduce an IR brane, at $y = y_1<y_s$, responsible for electroweak breaking. The 5D action of the model, including the stabilizing bulk scalar $\phi(x,y)$, with mass dimension $3/2$, reads as
\begin{eqnarray}
S &=& \int d^5x \sqrt{|\det g_{MN}|} \left[ -\frac{1}{2\kappa^2} R + \frac{1}{2} g^{MN}(\partial_M \phi)(\partial_N \phi) - V(\phi) \right]\nonumber \\
&-& \sum_{\alpha} \int_{B_\alpha} d^4x \sqrt{|\det \bar g_{\mu\nu}|} \lambda_\alpha(\phi)  
 -\frac{1}{\kappa^2} \sum_{\alpha} \int_{B_\alpha} d^4x \sqrt{|\det \bar g_{\mu\nu}|} K_\alpha  \,, \label{eq:action}
\end{eqnarray}
where $\kappa^2=1/(2M^3)$, with $M$ being the 5D Planck scale, $V(\phi)$ and $\lambda_\alpha(\phi)$ are the bulk and brane potentials of the scalar field $\phi$, and the index $\alpha=0 \; (\alpha=1)$ refers to the UV (IR)  brane. We will assume a $\mathbb Z_2$ symmetry ($y\to -y$) across the UV brane, which translates into boundary conditions on the fields, while we will impose matching conditions for bulk fields across the IR brane. Note that the fifth dimension continues beyond the IR brane until the singularity. The IR brane is responsible for the generation of the IR scale $\sim \TeV$, and contains the brane Higgs potential which spontaneously breaks the electroweak symmetry, as we will see. In addition, we will assume the Higgs field to be propagating in the bulk and localized toward the IR brane in order to solve the hierarchy problem. 

The parameter $\kappa^2$, can be traded by the parameter $N$ in the holographic theory by the relation~\cite{Gubser:1999vj}
$
N^2\simeq \frac{8\pi^2\ell^3}{\kappa^2}\,,
$
where $\ell\equiv 1/k$ is a constant parameter of the order of the Planck length, which determines the value of the 5D curvature. 
The metric $g_{MN}$ is defined in proper coordinates by
\begin{eqnarray}
ds^2 &=&g_{MN}dx^M dx^N\equiv e^{-2A(y)} \eta_{\mu\nu} dx^\mu dx^\nu - dy^2 \,,  \label{eq:metric}  
\end{eqnarray}
so that in Eq.~(\ref{eq:action}) the 4D induced metric is $ \bar g_{\mu\nu}=e^{-2A(y)}\eta_{\mu\nu}$, where the Minkowski metric is given by $\eta_{\mu\nu} =\diag(1,-1,-1,-1)$. 
 The last term in Eq.~(\ref{eq:action}) is the usual Gibbons-Hawking-York (GHY) boundary term~\cite{York:1972sj,Gibbons:1976ue}, where $K_{\alpha}$ are the extrinsic UV and IR curvatures. In terms of the metric of Eq.~(\ref{eq:metric}) the extrinsic curvature term reads as~\cite{Megias:2018sxv} $K_{0,1} = \mp 4 A^\prime(y_{0,1})$.

The equations of motion (EoM) read then as~\footnote{From here on the prime symbol $(\,{}^\prime\,)$ will stand for the derivative of a function with respect to its argument, and the dot symbol $(\dt{\phantom{a}})$ derivative only with respect to the conformal coordinate $z$ related to $y$ by $dy=e^{-A}dz$.}
\begin{eqnarray}
&&A^{\prime\prime}
= \frac{\kappa^2}{3} \phi^{\prime \, 2} + \frac{\kappa^2}{3} \sum_\alpha \lambda_\alpha(\phi) \delta(y - y_\alpha)  \,, \label{eq:eom1}\\
&&A^{\prime\, 2} 
= -\frac{\kappa^2}{6} V(\phi) + \frac{\kappa^2}{12} \phi^{\prime\, 2} \,,  \label{eq:eom2}\\
&&\phi^{\prime\prime} - 4 A^\prime \phi^\prime = V^\prime(\phi) + \sum_\alpha \lambda_\alpha^\prime(\phi) \delta(y - y_\alpha)  \,. \label{eq:eom3}
\end{eqnarray}
The EoM in the bulk can also be written in terms of the superpotential $W(\phi)$ as~\cite{DeWolfe:1999cp}
\begin{equation}
\phi^\prime = \frac{1}{2} \frac{\partial W}{\partial \phi} \,, \qquad A^\prime = \frac{\kappa^2}{6} W \,, \label{eq:phiA}
\end{equation}
and
\begin{equation}
V(\phi) = \frac{1}{8} \left( \frac{\partial W}{\partial \phi} \right)^2 - \frac{\kappa^2}{6} W^2(\phi) \,. \label{eq:V}
\end{equation}
The localized terms impose the following constraints in the UV ($\alpha=0$) and IR ($\alpha = 1$) branes,
\begin{equation}
A^\prime(y) \Big|^{y_\alpha^+}_{y_\alpha^-} = \frac{\kappa^2}{3} \lambda_\alpha(\phi_\alpha) \,,  \qquad \phi^\prime(y) \Big|^{y_\alpha^+}_{y_\alpha^-}  = \frac{\partial\lambda_\alpha(\phi_\alpha)}{\partial\phi} \,, \label{eq:constraints}
\end{equation}
where $\phi_\alpha = \phi(y_\alpha)$ and $y_\alpha^\pm=y_\alpha\pm \epsilon$. As mentioned above, we will assume $\mathbb Z_2$ symmetry across the UV brane, and one finds the following boundary conditions in the UV brane
\begin{equation}
A^\prime(y_0) = \frac{\kappa^2}{6} \lambda_0(\phi_0) \,, \qquad \phi^\prime(y_0) = \frac{1}{2} \frac{\partial\lambda_0(\phi_0)}{\partial\phi} \,, \label{eq:BC_UV}
\end{equation}
while Eq.~(\ref{eq:constraints}) for the IR brane corresponds to the jumping conditions
\begin{equation}
\Delta A^\prime(y_1) = \frac{\kappa^2}{3} \lambda_1(\phi_1) \,,  \qquad \Delta \phi^\prime(y_1)  = \frac{\partial\lambda_1(\phi_1)}{\partial\phi} \,, \label{eq:matching}
\end{equation}
where $\Delta X(y_1) \equiv X(y_1^+) - X(y_1^-)$ is the function jump.

For concreteness we consider for the brane potentials the form
\be
\lambda_\alpha(\phi) = \lambda_\alpha(v_\alpha) + \lambda_\alpha^\prime(v_\alpha)  (\phi - v_\alpha)  + \frac{\gamma_\alpha}{2}(\phi-v_\alpha)^2
\label{eq:brane-potentials}
\ee
in the stiff limit where $\gamma_\alpha\to\infty$, which fixes the brane minima at $\phi_\alpha=v_\alpha$.
Then, the UV boundary conditions~(\ref{eq:BC_UV}) read
\begin{align}
\lambda_0(v_0) = W(\phi_0) \,, \qquad  \lambda_0^\prime(v_0) = W^\prime(\phi_0) \,.
\label{eq:BCWprime}
\end{align}
Regarding the IR brane, a way to have matching conditions compatible with the EoM is by assuming~$\lambda_1(v_1) = \lambda_1^\prime(v_1) = 0$, i.e. a brane potential of the form
\begin{equation}
\lambda_1(\phi) =   \frac{\gamma_1}{2}(\phi-v_1)^2 \,,
\end{equation}
which implies that $A^\prime(y)$ and $\phi^\prime(y)$ are continuous functions at $y = y_1$, cf.~Eq.~(\ref{eq:matching})~\footnote{A non-vanishing value for $\lambda_1(v_1)$ would lead to a jump $\Delta A^\prime(y_1)$, and this would also demand a jump in the superpotential, $\Delta W(\phi_1)$ (by Eq.~(\ref{eq:phiA})). But then $W^\prime(\phi_1)$ would be divergent, and consequently the jump $\Delta \phi^\prime(y_1)$ would also be divergent. On the other hand, if one assumed a non-vanishing value for $\lambda_1^\prime(v_1)$, and consequently a finite jump $\Delta \phi^\prime(y_1)$, this would also induce a jump  $\Delta W^\prime(\phi_1)$. In the following we will use analytical (and continuous) expressions for the superpotential, so that we will not address this possibility.}.

Let us now consider the following ansatz for the superpotential
\begin{equation}
W(\phi) = \frac{6k}{\kappa^2} (1 + e^{\nu \phi}) \,,
\end{equation}
where the parameter $\nu$ has mass dimension $-3/2$.
Then, using Eq.~(\ref{eq:V}), it follows that the scalar potential is
\begin{equation}
V(\phi) = -\frac{6 k^2}{\kappa^2} \left[ 1 + 2 e^{\nu\phi} + \left( 1 - \frac{3\nu^2}{4\kappa^2} \right) e^{2\nu\phi} \right] \,.
\end{equation}
The EoM~(\ref{eq:phiA}) can be solved analytically with this ansatz, and the solution reads
\begin{equation}
\phi(y) = -\frac{1}{\nu} \log\left[ \frac{3k\nu^2}{\kappa^2} (y_s - y) \right] \,, \quad A(y) = ky - \frac{\kappa^2}{3\nu^2} \log\left[1 - \frac{y}{y_s} \right]  \,, \label{eq:phiA_ana}
\end{equation}
where we have chosen $A(0) = 0$. There is a singularity at $y = y_s$, while near the UV boundary $y \ll y_s$ the geometry is AdS$_5$, $A(y)\simeq k y$. 

It is useful to define the metric also in conformally flat coordinates defined by
\begin{equation}
ds^2 = e^{-2A(z)} \left( \eta_{\mu\nu} dx^\mu dx^\nu - dz^2 \right) \,, \label{eq:conformal_coord}
\end{equation}
where $dz = e^{A(y)}dy$. One can easily find, for $\nu > 0$, that
\begin{equation}
\rho \cdot (z - z_0 ) = \Gamma[1-\kappa^2/(3\nu^2),k(y_s - y)] - \Gamma[1-\kappa^2/(3\nu^2),ky_s] \,,  \label{eq:z_analytical}
\end{equation} 
with
\begin{equation}
\rho = k(k y_s)^{-\kappa^2/(3\nu^2)} e^{-k y_s} \,,  \label{eq:rho}
\end{equation}
where $z_0 = 1/k$ corresponds to the location of the UV brane, and $\Gamma[a,x]$ is the upper incomplete gamma function. As we will see below, $\rho$ is a scale of the order of $\TeV$, so that it is the relevant mass scale for the 4D spectrum, and it is responsible for the mass gap in the spectrum.

Let us now discuss the behavior of the background solution, in the deep IR close to the singularity, i.e. $y \simeq y_s$. Then, the asymptotic behavior of $z(y)$ is
\begin{equation}
z(y) \propto \textrm{const} + (y_s - y)^{1-\kappa^2/(3\nu^2)} + \cdots \,.
\end{equation}
The second term in the right-hand side is zero (divergent) in the limit $y\to y_s$ for $\nu > \kappa/\sqrt{3}$ ($\nu < \kappa/\sqrt{3}$), and this means in the former case the existence of a singularity at some finite value of the conformal coordinate, i.e. $z_s < \infty$. 

\begin{figure}[htb]
\centering
\includegraphics[width=7cm]{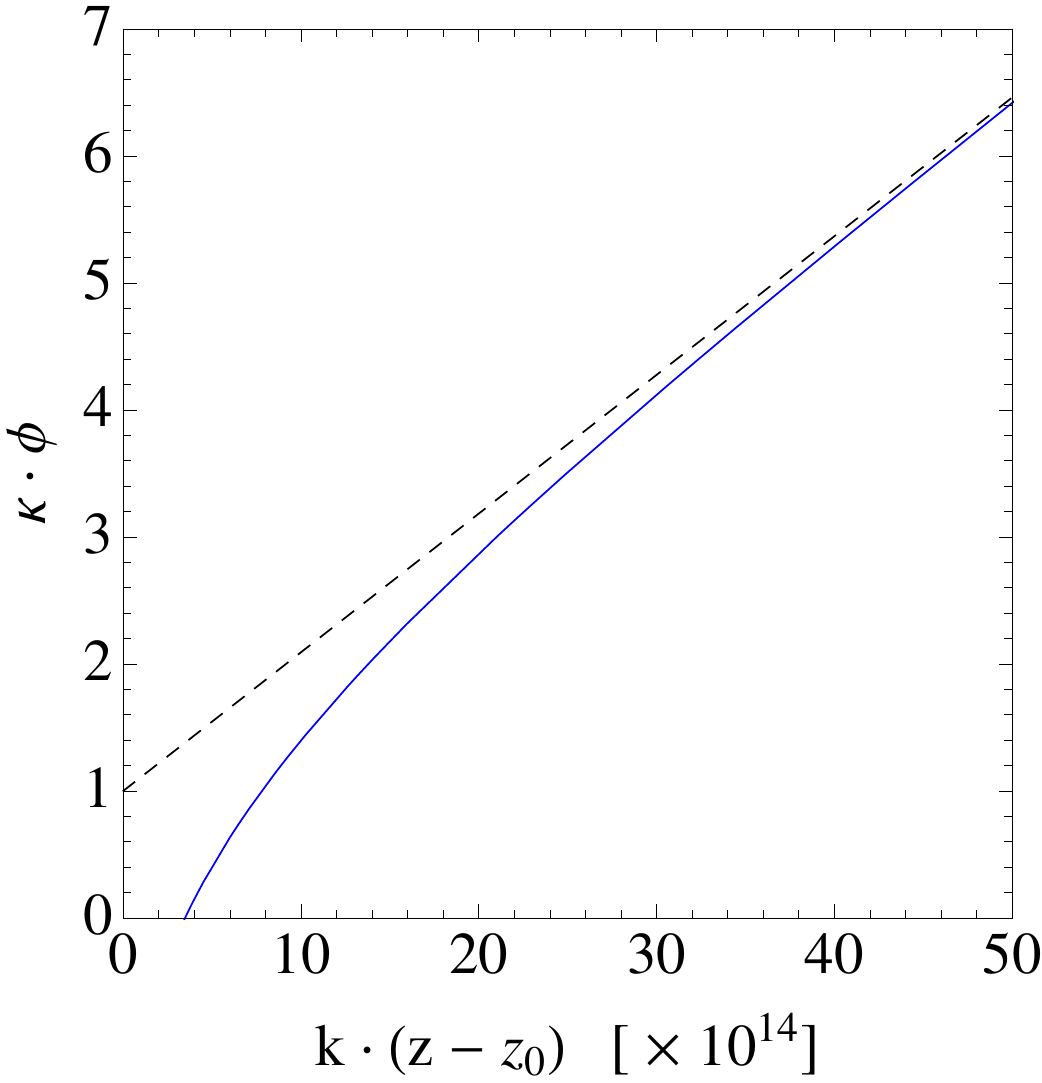} \hspace{0.5cm}
\includegraphics[width=7cm]{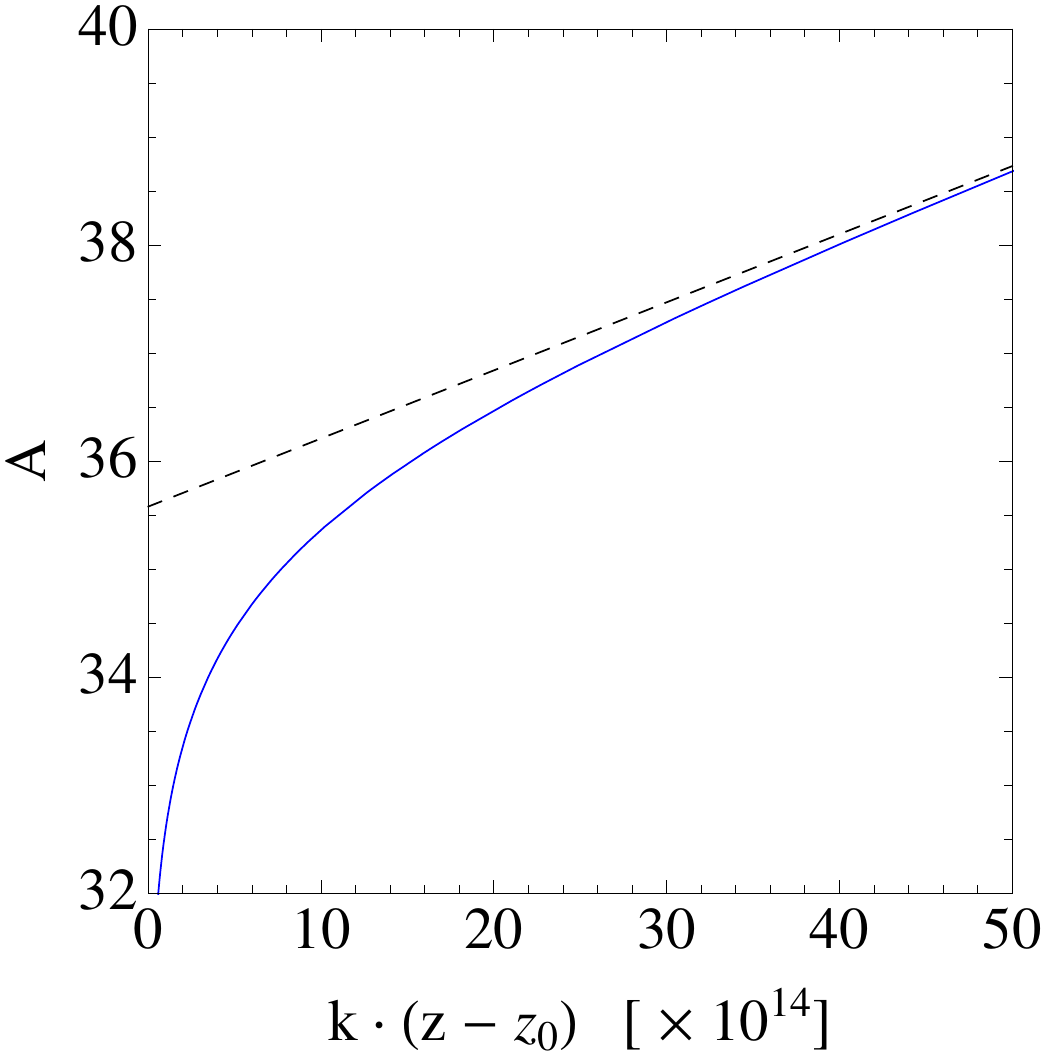}
\caption{\it Left panel: Scalar field $\phi$ as a function of the conformal coordinate $z$. Right panel: Warp factor of the metric $A$ as a function of $z$. We display the exact results in solid blue lines, and the linear behavior of Eqs.~(\ref{eq:phi_linear}) and (\ref{eq:A_linear}) in dashed lines. We have used $\nu = \kappa/\sqrt{3}$ and  $k y_s = 31.55$, as follows by assuming $A_1 = 35$ and $c=1$ (cf. Sec.~\ref{sec:Brane_stabilization}).
}
\label{fig:phiz}
\end{figure}

In the critical case $\nu = \kappa/\sqrt{3}$ the asymptotic behavior is 
\begin{equation}
\rho z = c_0 - \log(k(y_s - y)) + {\cal O}(y_s - y) \,,
\end{equation}
where~$c_0 = \rho z_0 - \gamma_E - \Gamma(0,k y_s)\simeq -\gamma_E$, with $\gamma_E$ being the Euler's constant. In this case the domain of the conformal coordinate is $z_0 \leq z < +\infty$. Moreover, the scalar field profile behaves in this case as
\begin{equation}
\phi(z) \simeq \frac{\sqrt{3}}{\kappa} \left( \rho z - c_0 \right)  + {\cal O}\left(e^{-\rho z}\right)     \,, \qquad \textrm{for} \qquad z \to +\infty \,, \label{eq:phi_linear}
\end{equation}   
and the warp factor of the metric
\begin{equation}
A(z) \simeq \rho z - c_0 - \log(\rho/k) +  {\cal O}\left(e^{-\rho z}\right)   \,, \qquad \textrm{for} \qquad z \to +\infty \,. \label{eq:A_linear}
\end{equation} 
In this way, both $\phi$ and $A$ behave linearly in terms of the conformal coordinate in the IR region.
As we will see in the next section, this is the only case that allows for the existence of a continuum KK spectrum. Notice that in this case, $\nu=\kappa/\sqrt{3}$, it is convenient to define
\begin{equation}
z_0 = \frac{\Gamma(0,k y_s)}{\rho} = \frac{1}{k} \left( 1 + {\cal O}((ky_s)^{-1}) \right) \,,
\end{equation}
as then Eq.~(\ref{eq:z_analytical}) has the property $ z(y) \stackrel[y \to -\infty]{\longrightarrow}{} 0 $. Plots of the dimensionless quantities, $\kappa \phi$ and $A$ as functions of the conformal coordinate are shown in Fig.~\ref{fig:phiz}.

\section{Brane and singularity stabilization}
\label{sec:Brane_stabilization}

We have assumed brane potentials $\lambda_\alpha(\phi)$ fixing the dilaton on the branes to specific values $\phi_\alpha=v_\alpha$. This fixing can stabilize the brane UV/IR distance, by the Goldberger-Wise mechanism, and fix the position of the singularity according to the solution of the hierarchy problem. We will do it in the critical case where $\nu=\kappa/\sqrt{3}$ for which
\begin{equation}
\phi(y) = -\frac{\sqrt{3}}{\kappa} \log\left[ k(y_s - y) \right] \,, \quad A(y) = ky -  \log\left[1 - \frac{y}{y_s} \right]  \,,\quad \rho\equiv k(ky_s)^{-1}e^{-ky_s} \,, \label{eq:phiA_final}
\end{equation}
where $\rho$ is the spectrum mass gap, as we will see, and seek for solutions with $y_1<y_s$. As the Higgs vacuum expectation value will be fixed at the IR brane, as we will see in Sec.~\ref{sec:Higgs}, we need to define the scale
\be
\bar\rho = k \, e^{-A(y_1)}
\ee
of the order of the TeV, so that we have to fix $A(y_1)\equiv A_1\simeq 35$ to solve the hierarchy problem.

Moreover, as both $\rho$ and $\bar\rho$ should be of the order of the TeV, we can impose the condition that
\be
\bar\rho=c\rho
\label{eq:c}
\ee
with $c=\mathcal O(1)$ constant. Then using the explicit expressions of $A(y)$ and $\rho$ (cf.~Eq.~(\ref{eq:phiA_final})), this condition leads to
\begin{equation}
ky_1 =k y_s -\cW(c) \,,
\end{equation}
where $\cW(x)$ is the Lambert function~\footnote{The Lambert function is the solution of the equation $c = \cW(c) e^{\cW(c)}$. For $c=0$ it vanishes while for $c>0$, $\cW(c)>0$. For instance, for $c=1$, the IR brane is located at $ky_1 \simeq ky_s -0.57$.}, so that $y_1$ is located before the singularity. 

Using this result, the values of the scalar field and warp factor in the IR brane turn out to be
\begin{equation}
v_1 = -\frac{\sqrt{3}}{\kappa}\log \left( \cW(c) \right) \,, \quad A(y_1) = k y_s + \log\left( ky_s/c \right) \,, \label{eq:A1}
\end{equation}
while in the UV brane they are
\begin{equation}
v_0 = - \frac{\sqrt{3}}{\kappa} \log(k y_s) \,, \quad A(0) = 0 \,.
\end{equation}
\begin{figure}[htb]
\centering
\includegraphics[width=7cm]{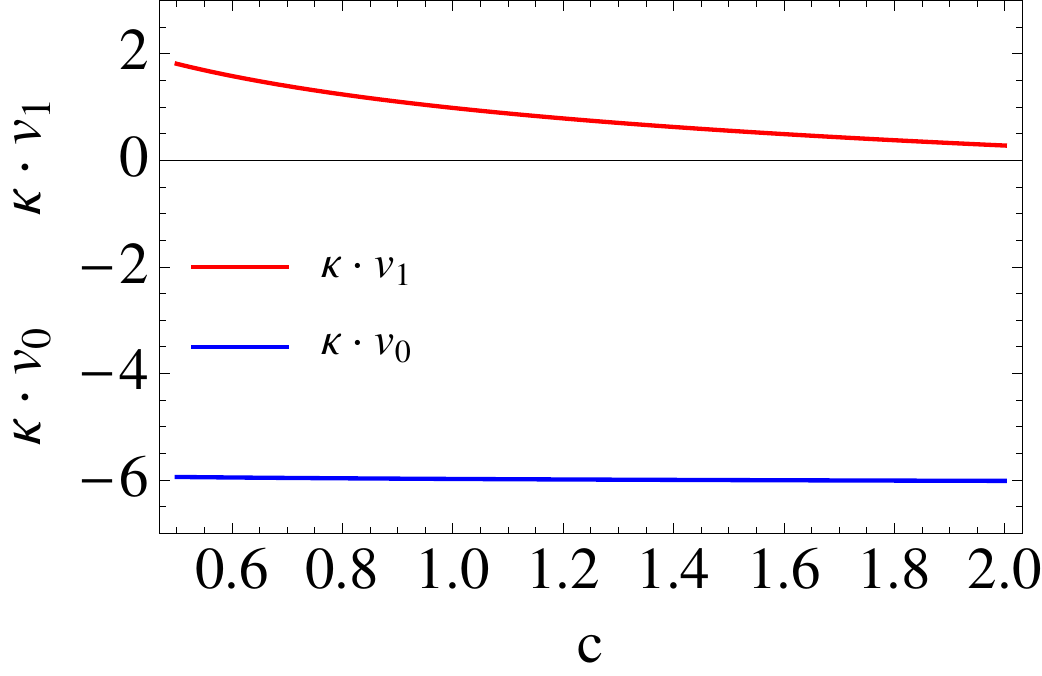} \hspace{1.0cm}
\includegraphics[width=7cm]{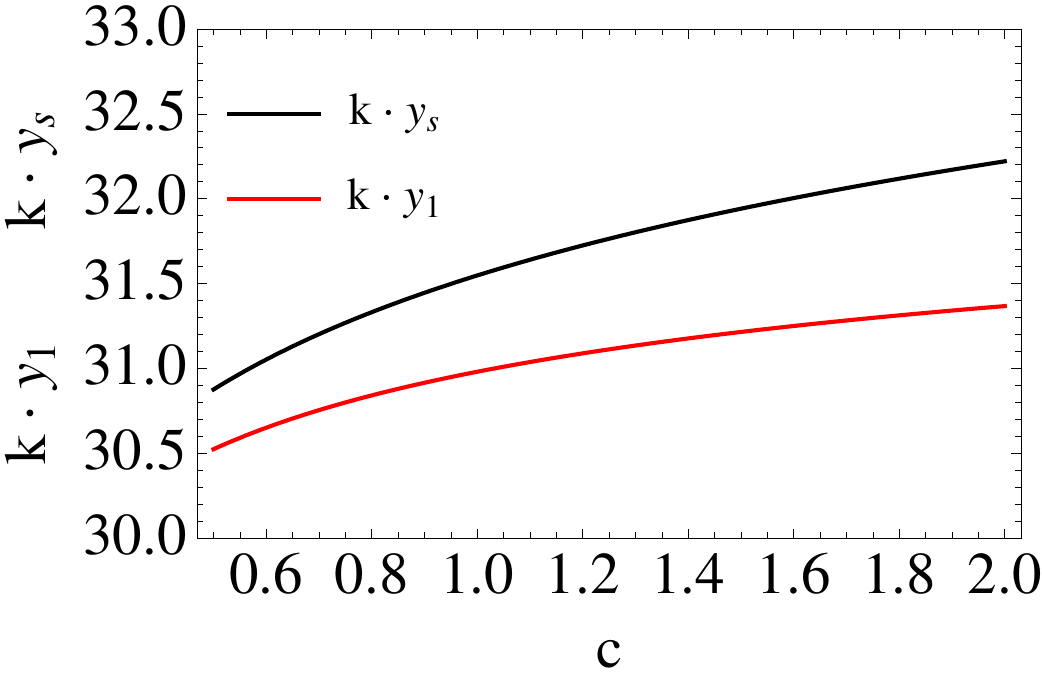} 
\caption{\it Left panel: $\kappa v_0$ and $\kappa v_1$ as a function of the parameter $c$. Right panel: Location of the IR brane, $ky_1$, and singularity, $ky_s$, as a function of the parameter $c$. We have used  $A_1 = 35$ and $\nu = \kappa/\sqrt{3}$.
}
\label{fig:braneIR}
\end{figure} 
The solution of $y_s$ from the second equation of~(\ref{eq:A1}) leads to
\begin{equation}
ky_s =   \cW(c \, e^{A_1}) \,,
\end{equation}
and then one can express $y_1$ and $v_0$ also in the form
\begin{equation}
ky_1 = \cW(c \, e^{A_1}) - \cW(c) \,, \quad \textrm{and} \quad  v_0 = - \frac{\sqrt{3}}{\kappa} \log \left( \cW(c \, e^{A_1}) \right) \,. \label{eq:y1_phi0}
\end{equation}
The values of $\kappa v_0$ and $\kappa v_1$ are plotted in the left panel of Fig.~\ref{fig:braneIR} as functions of the parameter $c$ defined in Eq.~(\ref{eq:c}). We can see that, in units of $\kappa$, they are always in absolute value $\mathcal O(\textrm{few})$.

Finally, it is possible to obtain the explicit dependence of the parameter $c$ in terms of $y_1$ and $A_1$ by solving the first equation in~(\ref{eq:y1_phi0}). The result is
\begin{equation}
c = k y_1   \frac{\exp\left( \frac{e^{A_1} k y_1}{e^{A_1} - e^{k y_1}} \right)}{e^{A_1} - e^{k y_1}} \,.
\end{equation}
Plots of $k y_s$ and $k y_1$ as functions of the parameter $c$ are explicitly shown in the right panel of Fig.~\ref{fig:braneIR}.

\section{The Higgs sector}
\label{sec:Higgs}

We have previously introduced the IR brane, stabilized at a distance $y=y_1$, with the only purpose of triggering electroweak symmetry breaking. In the simplest theory where the Higgs is a 5D bulk doublet 
\begin{equation}
H(x,y) = \frac{1}{\sqrt{2}} e^{i\chi(x,y)}  \left(
{ \begin{array}{c}
  0 \\
  h(y) + \widehat H(x,y) \\
  \end{array}
} \right) \,,
\end{equation} 
with an action given by
\begin{equation}
S_5 = \int d^5x\sqrt{|\det g_{MN}|} \left[ |D_M H|^2 - V(H) \right]-\int d^4x\sqrt{-g_{ind}}(-1)^\alpha \lambda^\alpha(H)\delta(y-y_\alpha) \,,
\label{eq:S5Higgs}
\end{equation}
where $V(H)=M^2(\phi)|H|^2$ is the 5D Higgs potential, electroweak breaking is triggered by the brane potentials defined as
\be
\lambda^0(H)=2M_0|H|^2,\quad \lambda^1(H)=M_1|H|^2-\gamma|H|^4\ .
\ee
Here the mass dimension of the Higgs field is $3/2$ and that of $\gamma$ is $-2$.

The background Higgs field is then determined from the EoM~\cite{Cabrer:2011fb,Megias:2015ory}
\begin{equation}
  h^{\prime\prime}(y) - 4 A^\prime(y) h^\prime(y) - \frac{\partial V}{\partial h} = 0\,, 
\label{eq:h}
\end{equation}
with the boundary conditions on the UV brane
\be
h'(0)=\left.\frac{1}{2}\frac{\partial \lambda^0(h)}{\partial h}\right|_{y=0} \,,
\label{eq:UVBCh}
\ee
and the jump conditions on the IR brane
\be
\Delta h^\prime(y_1)=-\left.\frac{\partial\lambda^1(h)}{\partial h}\right|_{y_1} \,.
\label{eq:IRBCh}
\ee
As we want to have a Higgs profile  $h(y)\propto e^{a ky}$, we can simply define a bulk Higgs potential with $M^2(\phi)=a k(a k-2\kappa^2 W(\phi)/3)$.
In this case the general solution to Eq.~(\ref{eq:h}) is given by
\be
h(y)=e^{a ky}\left[c_1+c_2 k\int^y_0 e^{4 A(y')-2a ky'}dy'
\right] \,,
\label{eq:solH}
\ee
where $c_1$ and $c_2$ are integration constants, to be determined from boundary and jump conditions. We will consider the solution (\ref{eq:solH}) in two regions: region A ($0<y\leq y_1$), and region B ($y_1<y<y_s$), with four integration constants: $c_1^A,\, c_2^A,\, c_1^{B},\, c_2^{B}$. 

\begin{figure}[htb]
\centering
\includegraphics[width=7cm]{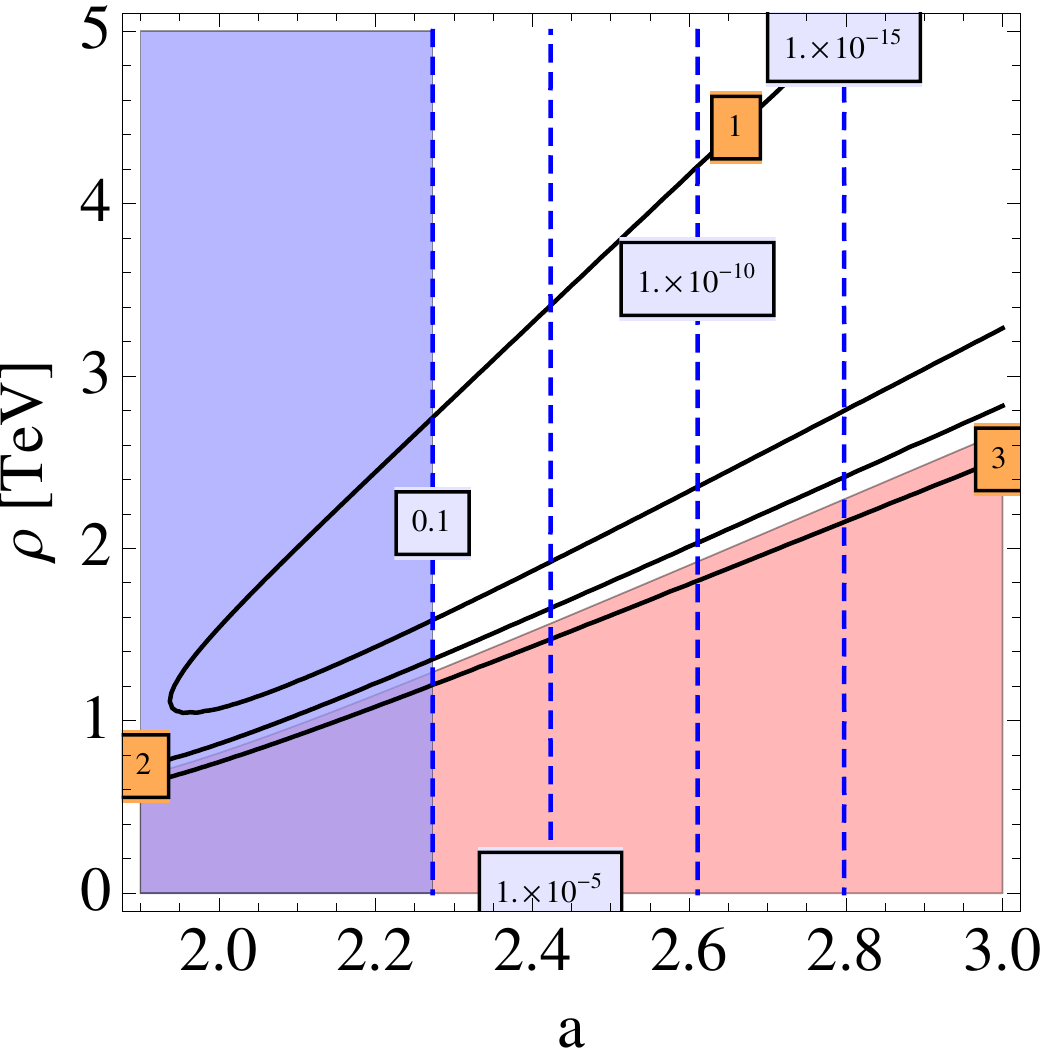} \hspace{1.0cm}
\includegraphics[width=7.0cm]{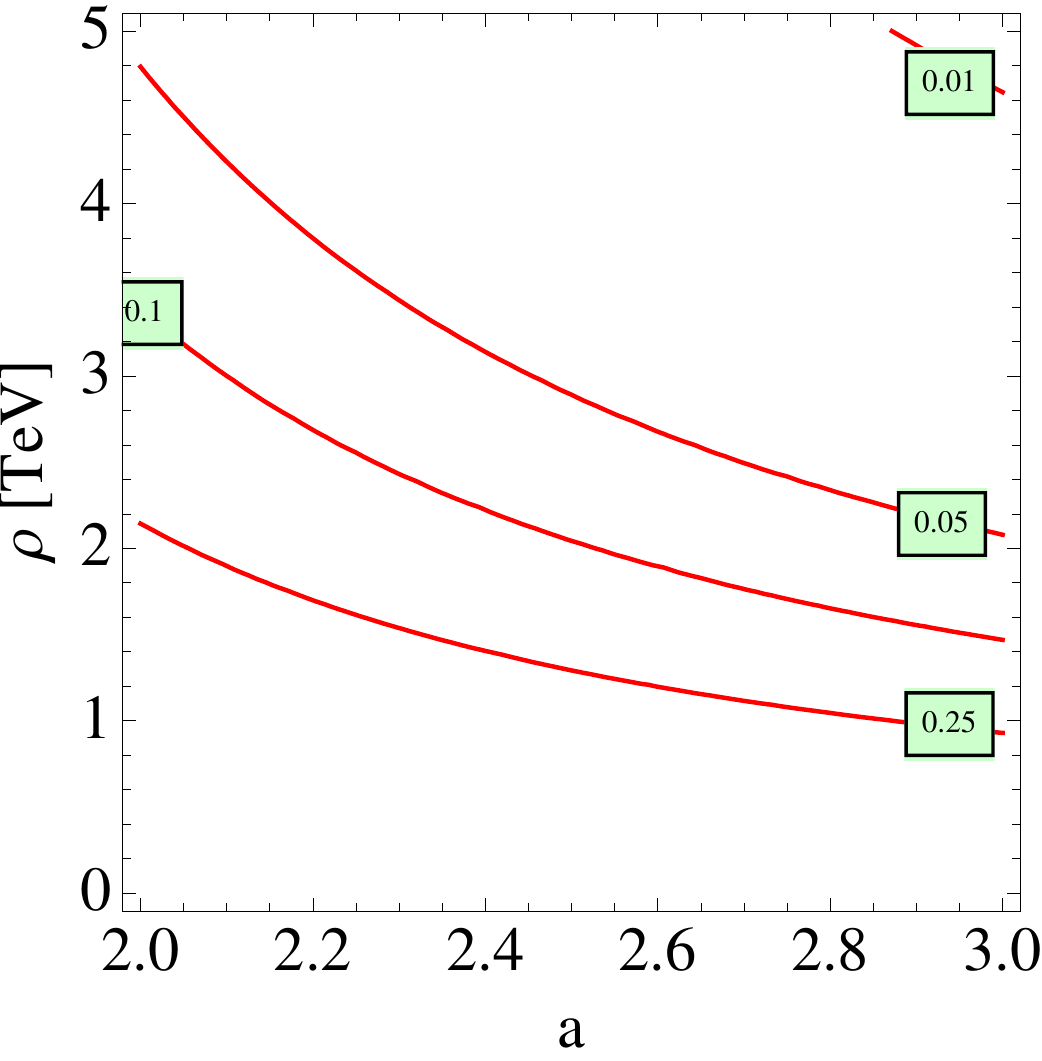}
\caption{\it Left panel: Dashed (blue) lines are contour plots of $F(y_1)$. Solid (black) lines provide the $\chi^2$ allowed region by the oblique $S$ and $T$ parameters at 1, 2 and 3 $\sigma$. We display in blue the region in which $F(y_1) > 0.1$, and in red the excluded region for oblique parameters at $95\%$ C.L. Right panel: Contour lines of the parameter $M_1/k$ as given from Eq.~(\ref{eq:Higgsmass}), where $c=1/4$ has been taken.
}
\label{fig:plotEugenio}
\end{figure}

We will impose on region A the boundary and jump conditions, Eqs.~(\ref{eq:UVBCh}) and (\ref{eq:IRBCh}), corresponding to the UV and IR branes, respectively. In particular the boundary condition on the UV brane imposes the condition $c_2^A = (M_0/k-a) \cdot c_1^A$, and the solution in region A is written as
\be
h_A(y)=e^{a ky}c_1^A\left[ 1-(M_0/k-a)F(0)+(M_0/k-a)F(y)\right],
\ee
where the function $F(y)$, as defined by $F'(y)=ke^{4A(y)-2a ky}$, can be obtained analytically as
\be
F(y) = - (2(a-2))^3 (ky_s)^4 e^{-2(a-2)ky_s} \, \cdot \Re  \Gamma\left[-3,-2(a-2)k(y_s-y) \right]  \,,  \label{eq:F}
\ee
where $\Gamma[n,x]$ is the upper incomplete gamma function, and $\Re$ stands for the real part. The approximate values of $F(y)$ at $y=0$ and $y=y_1$ are, respectively,
\begin{equation}
F(0) \simeq -\frac{1}{2(a-2)}  \,, \qquad  F(y_1) \simeq \frac{(k y_s)^4}{3 \cW(c)^3 } e^{-2(a-2)ky_s}  \,.
\end{equation}
As pointed out in Ref.~\cite{Cabrer:2011fb}, to keep the exponential solution without the need of the fine-tuning $M_0=a k$, we must require the function $F(y)$ to be small. Since $F$ is a monotonically increasing function of $y$, it will be enough to guarantee that $F(y_1)\ll 1$. The structure of the function $F$ implies that a necessary condition is that $a>2$. Contour lines of $F(y_1)$ are shown in the left panel of Fig.~\ref{fig:plotEugenio} (dashed lines).

Finally we have to impose the IR jump and continuity conditions 
\be
\Delta h^\prime(y_1)=h'_B(y_1)-h'_A(y_1)=-[M_1-\gamma h^2(y_1)]h(y_1),\quad h_A(y_1)=h_B(y_1) \,,
\label{eq:IRjump}
\ee
as well as the regularity condition at $y=y_s$ in region B, which fixes $c_2^B=0$, as $F(y)$ is singular at $y_s$. After imposing regularity, the jump conditions in Eq.~(\ref{eq:IRjump}) are satisfied by fixing the integration constants $c_1^A$ and $c_1^B$ as
\be
c_1^B=[1-(M_0/k-a)F(0)] \cdot c_1^A= \sqrt{M_1/\gamma} \cdot e^{-a ky_1} \,,
\ee
by which the continuity of $h^\prime(y_1)$ follows.

As for the profile of the zero mode of the Higgs excitation, $\widehat{H}(y)$, for $m_H=0$ it satisfies the same equation as the background, so that $\widehat{H}(y)\simeq h(y)$. This approximation also holds for a light Higgs, so that in particular it should be a good approximation for the physical Higgs mass $m_H= 125$ GeV, as was shown in detail in Ref.~\cite{Cabrer:2011fb}. Of course, as we will see in Sec.~\ref{sec:continuum_spectrum}, for momenta $p>3\rho/2$ there is a continuum of states, as for the other fields in the model.

The next step is to impose correct electroweak symmetry breaking by using the expression, for $v=246$ GeV~\cite{Cabrer:2011fb},
\be
v^2=\int_0^{y_s} dy \, h^2 e^{-2 A} \,.
\ee
After using our metric $A(y)$, as well as the equations obtained in this section, one can get the expression for the dimensionless parameter $M_1/(k^3\gamma)$, as
\be
\frac{M_1}{k^3\gamma}=\frac{v^2}{\rho^2}\cdot f(a,c),\quad \textrm{with  } f(a,c)= \frac{4(a-1)^3}{e^{2a \mathcal W(c)}}\,.
\label{eq:M1gamma}
\ee

Moreover, by constructing the effective 4D theory, and using the IR Higgs brane potential, one can get the expression for the Higgs mass, as was done in Ref.~\cite{Cabrer:2011fb}. The result for the SM Higgs potential $V_{SM}=-\mu^2|H_{SM}|^2+\lambda |H_{SM}|^4$, where
\be
H(x,y)\equiv 2(a-1)^{3/2}\sqrt{k} \,\frac{h(y)}{h(y_s)}\,\frac{k}{\rho} H_{SM}
\ee
is
\be
\mu^2=(M_1/k)f(a,c) c^4 \rho^2,\quad \lambda=f^2(a,c)c^4  k^2\gamma \,,
\ee
and therefore the Higgs mass
\be
m_H^2=2 c^4 f(a,c)M_1 \rho^2/k\,,
\label{eq:Higgsmass}
\ee
from where we can see that the ``natural" value of $m_H^2$ would be $\rho^2$, which triggers a little naturalness problem, as usually $\rho\gg m_H$. Therefore we have to tune the mass $M_1$ to small values, and this provides a measure of the size of the ``unnaturalness" in this class of theories. Contour lines of the parameter $M_1/k$ are shown in the right panel of Fig.~\ref{fig:plotEugenio}, in the plane $(a,\rho)$ for $c=1/4$. We can see that the tuning is typically $\mathcal O(10\%)$ for $\rho\lesssim 3$ TeV. Clearly the tuning is tougher for larger values of the parameter $c$, as it grows as~$c^4$.

We can now compute the $S$ and $T$ parameters along the lines of Ref.~\cite{Cabrer:2011fb}. The experimental bounds on the $S$ and $T$ parameters are given by~\cite{Olive:2016xmw}
\begin{equation}
S = 0.02 \pm 0.07 \,, \qquad T = 0.06 \pm 0.06 \,, \qquad  r \simeq 0.92 \,,
\end{equation}
where $r$ is the correlation. The results can be obtained analytically, although the expressions are rather cumbersome. The result is plotted in the left panel of Fig.~\ref{fig:plotEugenio} where the solid lines correspond to 1, 2 and 3~$\sigma$. We have used $c=1$, but the figure does not change appreciably when considering other values of $c$ in the interval $1/4 < c < 2$.

\section{Holographic Green's functions}
\label{sec:continuum_spectrum}

We will discuss in this section the holographic spectral functions, and UV-brane-to-UV-brane Green's functions, for the case of continuum spectra of KK modes. We will analyze separately the cases of KK gauge bosons, fermions, graviton, radion and the Higgs boson.

\subsection{Massless gauge bosons}
\label{subsec:gauge_bosons}

In the case of massless gauge bosons $A_\mu$ (i.e.~the SM photon and gluon) the Lagrangian is~\footnote{We are using in this section the gauge $A_5=0$.}
\be
\mathcal L= \int_0^{y_s} dy\left[ -\frac{1}{4}F_{\mu\nu}F^{\mu\nu}-\frac{1}{2}e^{-2A}A'_\mu A'_\mu  \right]\,.
\label{eq:Lagrangian_GB}
\ee
Defining $A_\mu(x,y)=f_A(y) A_\mu(x)$,
the EoM of the fluctuations is given by~\cite{Cabrer:2010si}
\begin{equation}
p^2 f_A + \frac{d}{dy}(e^{-2A} f_A^\prime(y)) = 0 \,,  \label{eq:fAy}
\end{equation}
where we have replaced the eigenvalue $m^2$ by $p^2$. In conformal coordinates, and after rescaling the field by ${ f}_A(z)= e^{A(z)/2}\hat f_A(z)$, we obtain the Schr\"odinger like form for the equation of motion
\begin{equation}
-\ddt{{\hat{f}}}_A(z) + V_A(z)  {\hat f_A}(z) = p^2 {\hat f_A}(z) \,, \label{eq:ftA}
\end{equation}
where the potential is
\begin{equation}
V_A(z) =  \frac{1}{4} {\dt A}^2(z) - \frac{1}{2} {\ddt A}(z)  \,. \label{eq:VA}
\end{equation}
An equivalent expression for the potential, in terms of the superpotential $W[\phi]$, is
\begin{eqnarray}
V_A(z) =  \frac{\kappa^2}{48} e^{-2A(z)} \left( \kappa^2 W^2[\phi(z)] - 2 (W^\prime[\phi(z)])^2  \right) \,.
\end{eqnarray}
This expression is valid for any $W[\phi]$. Close to the singularity, the potential behaves as 
\begin{equation}
V_A(z) \stackrel[z \to \infty]{\longrightarrow}{} \frac{1}{4} \rho^2  \,.
\end{equation}
This behavior can be easily understood from the property that, in this limit, ${\dt A}(z) \to \rho$ and ${\ddt A}(z) \to 0$. Then, we find the existence of a mass gap of the potential.

To compute the spectral density and Green's function we will use the holographic method. After Fourier transforming the coordinates $x^\mu$ into momenta $p^\mu$, we define~\footnote{For the sake of notational simplicity, we are using the same notation for functions and their Fourier transforms.}
\be
A_\mu(p,z)=f_A(p,z)a_\mu^{(4)}(p) \,,
\ee
where $f_A(p,z)$ satisfies Eq.~(\ref{eq:fAy}) and the 4D wave function $a_\mu^{(4)}(x)$ satisfies the 4D EoM
\be
[\eta^{\mu\nu}\Box -\partial^\mu\partial^\nu(1-1/\xi)+p^2\eta^{\mu\nu}]a_\nu^{(4)}(x)=0\,,
\ee
where we have considered the gauge fixing term $\mathcal L_{GF}=-1/(2\xi) [\partial^\mu a^{(4)}_\mu]^2$. Using now the EoMs into the Lagrangian, we can write the holographic Lagrangian as
\be
\mathcal L_{hol}=\frac{1}{2}e^{-A(z_0)}f_A(p,z_0) \dt{f}_A(p,z_0)a_\mu^{(4)}(p)P^{\mu\nu}a_\nu^{(4)}(p)\,,
\ee
where $P^{\mu\nu}(\xi)=\eta^{\mu\nu}-(1-1/\xi)p^\mu p^\nu/p^2$. If we fix the boundary condition at the UV brane as
\be
A_\mu(p,z_0)=a_\mu^0(p) \,,
\ee
where $a_\mu^0$ is the source coupled to the CFT vector operator $\mathcal J_\mu^A$, the holographic Lagrangian turns out to be (we have normalized the metric as $A(z_0)=0$)
\be
\mathcal L_{hol}=\frac{1}{2}\frac{\dt{f}_A(p,z_0)}{f_A(p,z_0)}\,a_\mu^{0}P^{\mu\nu}(\xi)a_\nu^{0} \,.
\label{eq:uno}
\ee
The two-point function is now the inverse of the bilinear operator in (\ref{eq:uno}) as
\be
G^{\mu\nu}_A(z_0,z_0;p)=[\eta^{\mu\nu}-(1-\xi)p^\mu p^\nu/p^2]G_A(z_0,z_0;p) \,,
\label{eq:correlator_gauge}
\ee
where the 4D Green's function $G_A(z_0,z_0;p)$ is
\be
G_A(z_0,z_0;p)=\frac{f_A(p,z_0)}{\dt f_A(p,z_0)}\equiv \int_0^\infty ds\frac{\sigma_A(s)}{s-p^2+i\epsilon} \,,
\ee
and the spectral density $\sigma_A$ is then obtained as
\be
\sigma_A(z_0,z_0;p)=\frac{1}{\pi}\, \textrm{Im}\left[\frac{f_A(p,z_0)}{\dt{f}_A(p,z_0)}\right] \,.
\ee

\begin{figure}[htb]
\centering
\includegraphics[width=7.5cm]{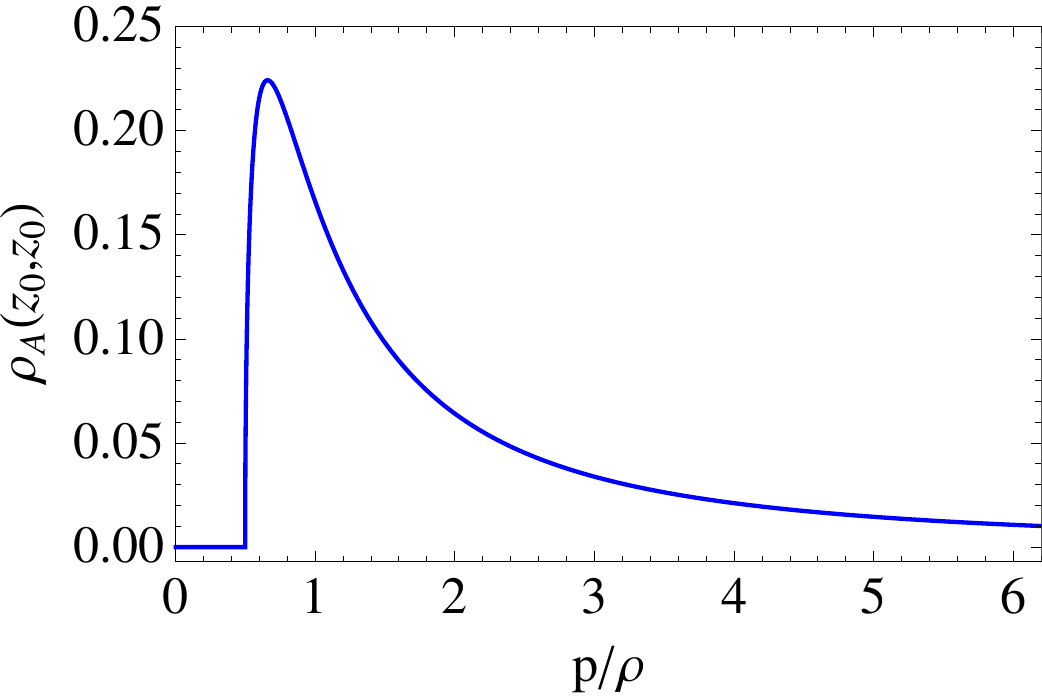} \hspace{0.25cm}
\includegraphics[width=7.5cm]{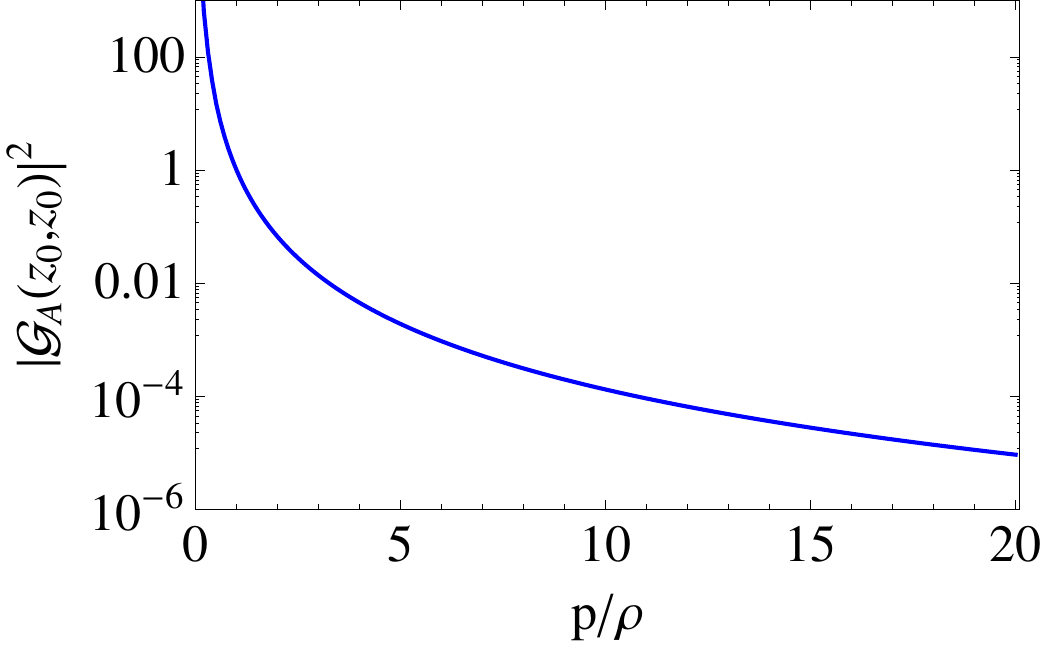}
\caption{\it Left  panel: Spectral density $\rho_A(z_0,z_0;p)$ for a continuum gauge boson.  Right panel: Absolute value squared of the Green's function, $\mathcal G_A(z_0,z_0;p)$, for gauge bosons. We have used $A_1 = 35$ and $c=1$ in both panels.}
\label{fig:spectral_gaugebosons}
\end{figure} 

The solution of Eq.~(\ref{eq:ftA}) in the IR is of the form
\begin{equation}
{\hat f}_A(z) \simeq  c_- e^{-\Delta z} + c_+ e^{\Delta z} \,,  \label{eq:fA_IR}
\end{equation}
with~$\Delta = \frac{\rho}{2}\sqrt{1- (2p/\rho)^2}$. The computation of the retarded Green's function demands the use of ``IR regular'' solutions for Euclidean AdS, i.e. the solution with $c_+=0$. This corresponds to outgoing wave boundary conditions after analytical continuation~\cite{Son:2002sd}. We have solved numerically Eq.~(\ref{eq:fAy}) by using the IR boundary condition mentioned above. 
As scale invariance is explicitly broken by the scale $\rho$, we have looked for a rescaling of the Green's function $G_A$ which makes it scale invariant (i.e.~invariant with respect to variation of the parameter $\rho$). The required rescaled Green's function
\be
 {\mathcal G}_A(z_0,z_0;p) \equiv (\rho^2/k) \mathcal W(k/\rho) G_A (z_0,z_0;p)
\label{eq:G_scaled}
\ee
is shown in Fig.~\ref{fig:spectral_gaugebosons}, where we plot, in the left panel, the corresponding scale invariant spectral function,
\be
\rho_A(z_0,z_0;p)\equiv (\rho^2/k)\mathcal W^2(k/\rho)\sigma_A(z_0,z_0;p)
\label{eq:rho_scaled}
\ee
and, in the right panel, the squared scale invariant Green's function $|\mathcal G_A(z_0,z_0;p)|^2$, as functions of $p/\rho$. The latter translates into the behavior $G_A(z_0,z_0;p) \sim 1/p^2$ for $p\gg \rho$. Moreover the Green's function has, on top of the continuum for momenta larger than the value of the mass gap, an isolated massless mode, which signals the contribution from the SM massless gauge boson, say the photon or the gluon. 

Notice that the scaling behaviors of the Green's function $G_A$, Eq.~(\ref{eq:G_scaled}), and its spectral density $\sigma_A$, Eq.~(\ref{eq:rho_scaled}), are different from each other. This situation can happen when the imaginary part of the Green's function is much smaller than its real part, in which case the absolute value of the Green's function is dominated by the real part, which then provides its global scaling. We will find this kind of behaviors for the Green's function and its spectral density for other fields in this paper.

In this way the correlator of the CFT field $\mathcal J_\mu^A$ is given by~\cite{PerezVictoria:2001pa} 
\be
\Delta_{\mu\nu}^A=\frac{\delta^2 S}{\delta a_\mu^0\delta a_\nu^0 }=\langle\mathcal J_\mu^A \mathcal J_\nu^A \rangle=\frac{P^{\mu\nu}(\xi)}{G_A(p^2)}\,,
\ee 
and the comparison with the correlator in a pure CFT~\cite{Grinstein:2008qk} of a vector $\mathcal J_\mu^A$ with anomalous dimension $d$,
\be
\Delta_{\mu\nu}\propto (-p^2-i\epsilon)^{d-2}\left[\eta_{\mu\nu}-\frac{2(d-2)}{d-1}p_\mu p_\nu/p^2\right]\,,
\label{eq:correlator_gaugeCFT}
\ee
yields (in the physical unitary gauge $\xi\to\infty$), as dimension of the operator $\mathcal J_\mu^A$, $d_A=3$. Of course, as the correlator (\ref{eq:correlator_gauge}) is not written in a pure conformal theory, one cannot really interpret $d_A$ as a true dimension of the operator $\mathcal J^A_\mu$.

\subsection{Massive gauge bosons}
\label{subsec:massive_gauge_bosons}

In the case of massive gauge bosons $A_\mu$ (i.e.~the SM $W_\mu$ and $Z_\mu$) there is an extra term in the 5D Lagrangian, Eq.~(\ref{eq:Lagrangian_GB}), as~\cite{Cabrer:2011fb}
\be
\Delta \mathcal L_{5}= -\frac{1}{2}M_A^2(y) A_\mu^2
\ee 
where~\footnote{For a nearly-constant profile $f_A$ this leads to $$m_A^2\simeq\frac{1}{y_s}\int_0^{y_s}M_A^2(y)dy\ .$$}
\be
M_A^2(y)=m_A^2\, y_s\, \frac{e^{2 a k y-2A(y)}}{\int_0^{y_s} e^{2 a k y-2 A(y)}dy} 
\ee
with $m_A=m_{W,Z}$ being the physical value of the gauge boson mass, and
which leads, after the decomposition $A_\mu(x,y)=f_A(y) A_\mu(x)$, to a modification of the EoM~(\ref{eq:fAy}) as
\be
p^2 f_A+\frac{d}{dy}\left(e^{-2A} f^\prime_A(y)\right)=M_A^2(y) f_A\ .
\ee
After changing to conformal coordinates and making the field replacement $f_A(z)=e^{A(z)/2}\hat f_A(z)$, we obtain the Schr\"odinger equation~(\ref{eq:ftA}) with a potential $V_A(z)+\Delta V_A(z)$, where $V_A(z)$ is the potential given in Eq.~(\ref{eq:VA}), and
\be
\Delta V_A(z)=M_A^2(y(z))\ .
\ee 
As $\lim_{z\to\infty} \Delta V_A(z)=0$, the mass gap is the same as that of the massless case, i.e.~$m_g=\rho/2$.  

\begin{figure}[htb]
\centering
\vspace{0.15cm}
\includegraphics[width=7.9cm]{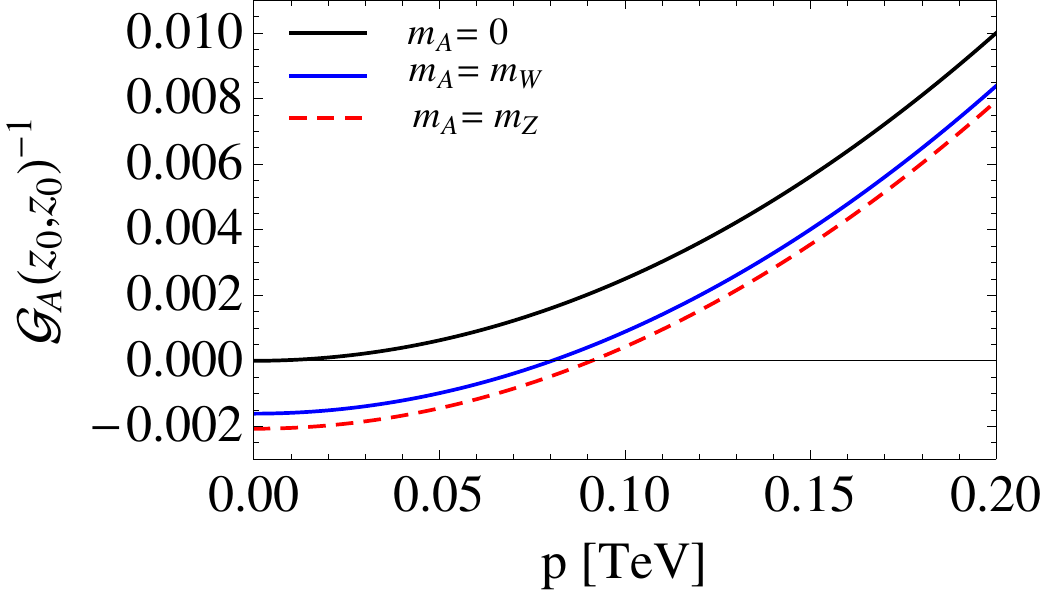} %\hspace{0.25cm}
\includegraphics[width=7.5cm]{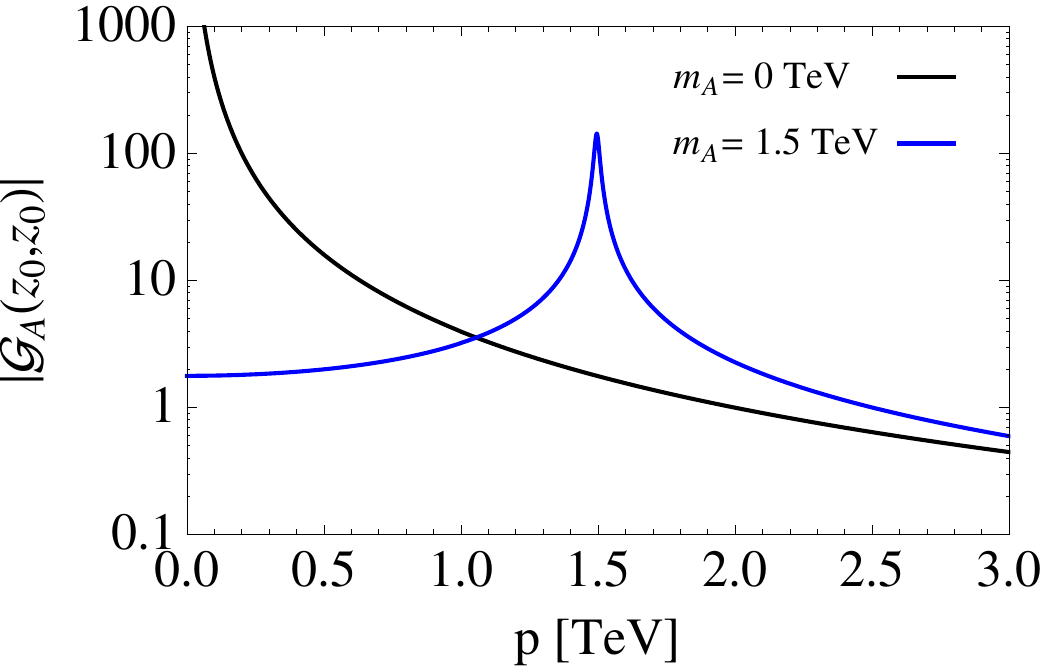} 
\caption{\it Left  panel: Inverse Green's functions for $a=2.4$ and different values of $m_A=0, m_W, m_Z$ as a function of $p$. Right panel: Absolute value of the Green's function, $\mathcal G_A(z_0,z_0;p)$, for an extra heavy gauge boson with mass $m_A=1.5$ TeV and $a=0$.  We have used $A_1 = 35$, $c=1$ and $\rho=2$ TeV in both panels.}
\label{fig:massive_gaugebosons}
\end{figure}

The computation of the holographic spectral and Green's functions follows the same lines as in Sec.~\ref{subsec:gauge_bosons}. In particular the spectral function shows a continuum for momenta $p\geq m_g$ and a Dirac delta function for $p=m_A$, which corresponds to a stable resonance at the order we are computing~\footnote{A non-zero width should be induced at the loop level, corresponding to the available decay channels of the massive gauge boson, as in the SM.}. This pole behavior should also show up in the Green's function $\mathcal G_A(z_0,z_0;p)$, which translates into a zero in the inverse Green's function. In the left panel of Fig.~\ref{fig:massive_gaugebosons} the inverse Green's
function, as a function of $p$, is shown for several values of $m_A=(0, m_W, m_Z)$, and where we have fixed $\rho=2$ TeV. We can see that the zeroes of the inverse Green's functions appear at $p=m_A$. In the case of some extra heavy gauge boson, e.g.~an extra $Z^\prime$ or $W^\prime$, with a mass $m_A>m_g$ the isolated resonance appears in the sea of the continuum. We have shown this hypothetical case in the right panel of Fig.~\ref{fig:massive_gaugebosons} where we have fixed $\rho=2$ TeV and $m_A=1.5$ TeV. As we can see the resonance gets a non-zero width and looks like a Breit-Wigner resonance. This phenomenon can be interpreted at this level as representing the propagator of the isolated resonance getting an imaginary part, and therefore the particle becoming unstable against decays into the lighter continuum KK resonances. This phenomenon was already studied in the case of unparticles in Ref.~\cite{Delgado:2008gj} with similar conclusions.

\subsection{Fermions}
\label{subsec:fermions}

In the case of fermions $\psi=(\psi_L,\psi_R)^T$, after rescaling the fields as $\psi_{L,R}(y) \to e^{2A}\psi_{L,R}(y)$, the action is~\cite{Cabrer:2011qb}
\be
\mathcal L=\int_0^{y_s}dy \left[e^A\, i\bar\psi /\hspace{-.22cm}\partial\psi-M_\psi(y)\bar\psi\psi + (\bar\psi_R\psi^\prime_L+\bar\psi_L^\prime\psi_R)
\right] \,,
\ee
where $M_\psi$ is a bulk mass of the fermions which in general will depend on $y$, in terms of which the equations of motion read as~\cite{Cabrer:2011qb}
\begin{equation}
m \psi_{L,R} = e^{-A}(M_\psi(y) \pm \partial_y) \psi_{R,L} \,,
\end{equation}
where we define the chiral fermions as $\psi_{L,R}=\frac{1}{2}(1\mp \gamma_5)\psi$ and $\gamma_5=\diag(-1,1)$. Using twice the equation in both versions, for left and right handed fields, we obtain the Schr\"odinger like form for the equations of motion (replacing $m^2$ by $p^2$)
\begin{equation}
-\ddt{\psi}_{L,R}(z) + V_{L,R}(z) {\psi}_{L,R}(z) = p^2 {\psi}_{L,R}(z) \,,
\end{equation}
where we are using conformal coordinates $z$, and the potential $V_{L,R}(z)$ is given by
\begin{equation}
V_{L,R}(z) = e^{-2A} M_\psi^2(z) + e^{-A}( \mp M_\psi(z) {\dt A}(z) \pm {\dt M_\psi}(z) ) \,. \label{eq:VLR}
\end{equation}
We will make the choice $M_\psi(z) = \epsilon c_\psi \frac{\kappa^2}{6} W(\phi(z))$ where $\epsilon = -1$ holds for 5D fermions with left-handed zero modes, while $\epsilon = +1 $ for 5D fermions with right-handed zero modes, and $c_\psi$ is an arbitrary constant. Then, this potential can be written as
\begin{eqnarray}
V_{L,R}(z) = \epsilon c_\psi \frac{\kappa^2}{36} e^{-2A(z)} \left( \kappa^2 (\epsilon c_\psi \mp 1) W^2[\phi(z)] \pm  3 (W^\prime[\phi(z)])^2  \right) \,.  \label{eq:VLRW}
\end{eqnarray}
Finally, it follows that the potentials have a mass gap, and its value is given by
\begin{equation}
V_{L,R}(z)  \stackrel[z \to \infty]{\longrightarrow}{}  \left( c_\psi \rho \right)^2 \,.
\end{equation} 

To compute the Green's function and spectral density we will again use holographic methods. We will define 
\be
\psi_{L,R}(p,z)=f_{L,R}(p,z) \psi^{(4)}_{L,R}(p) \,,
\ee
where $\psi^{(4)}_{L,R}(x)$ are the 4D plane-wave spinors satisfying the 4D Dirac equation with mass $p=\sqrt{p^2}$
\begin{equation}
i\bar\sigma^\mu\partial_\mu \psi_L^{(4)}(x)=p\psi_R^{(4)}(x) \,, \qquad
i\sigma^\mu\partial_\mu \psi_R^{(4)}(x)=p\psi_L^{(4)}(x) \,.
\end{equation}
 After using the EoM, the bulk action reduces to a pure boundary term on the UV brane~\cite{Cacciapaglia:2008ns}, which yields the holographic Lagrangian
 \be
 \mathcal L_{hol}=-\bar\psi_L(p,z_0)\psi_R(p,z_0)=-\bar f_L(p,z_0)f_R(p,z_0)\bar\psi_L^{(4)}(p)\psi_R^{(4)}(p) \,.
 \label{eq:Lholografico}
 \ee

 We now fix the boundary condition on the UV brane by fixing one of the component spinors, e.g.~the left handed one, to be the spinor $\psi_L^0$, which plays the role of a left-handed source coupled to the right-handed CFT operator $\mathcal O_R$, $\bar\psi^0_L\mathcal O_R+\bar{\mathcal O}_R\psi^0_L$, \textit{i.e.}
 \be
 \psi_L(p,z_0)= \psi_L^0(p) \,.
 \label{eq:fixingUV}
 \ee
Plugging now (\ref{eq:fixingUV}) into (\ref{eq:Lholografico}) results in
 \be
 \mathcal L_{hol}=-\frac{f_R(p,z_0)}{f_L(p,z_0)}\frac{\bar\psi_L^0\bar\sigma^\mu p_\mu \psi^0_L}{p}\,.
\label{eq:L_fermions}
 \ee
The two-point function is now given by the inverse of the quadratic term in Eq.~(\ref{eq:L_fermions}) as
\be
S_L=\sigma^\mu p_\mu\, G_L(z_0,z_0;p) \,,
\ee
where the 4D Green's function $G_L$ is given by
\be
G_L(z_0,z_0;p)=-\frac{1}{p} \frac{f_L(p,z_0)}{f_R(p,z_0)} \,, \label{eq:GL}
\ee
and the spectral density is obtained as
\be
\sigma_L(z_0,z_0;p)=\frac{1}{\pi}\textrm{Im}\, G_L(z_0,z_0;p)= -\frac{1}{\pi}\frac{1}{p}\textrm{Im}\,\left[\frac{f_L(p,z_0)}{f_R(p,z_0)}\right] \,. \label{eq:sigma_L}
\ee
The same analysis can be done for the right-handed spinor fixed on the UV brane to the spinor $\psi_R^0$, which couples to a left-handed CFT operator $\mathcal O_L$, $\bar\psi^0_R\mathcal O_L+\bar{\mathcal O}_L\psi^0_R$. This corresponds to a different CFT, and the result is
\begin{eqnarray}
G_R(z_0,z_0;p) &=& -\frac{1}{p}\frac{\bar f_R(p,z_0)}{\bar f_L(p,z_0)} \,, \\
\sigma_R(z_0,z_0;p) &=& \frac{1}{\pi}\textrm{Im}\, G_R(z_0,z_0;p)= \frac{1}{\pi}\frac{1}{p}\textrm{Im}\,\left[\frac{f_R(p,z_0)}{f_L(p,z_0)}\right] \,. \label{eq:sigma_R}
\end{eqnarray}
. 
\begin{figure}[htb]
\centering
\includegraphics[width=7cm]{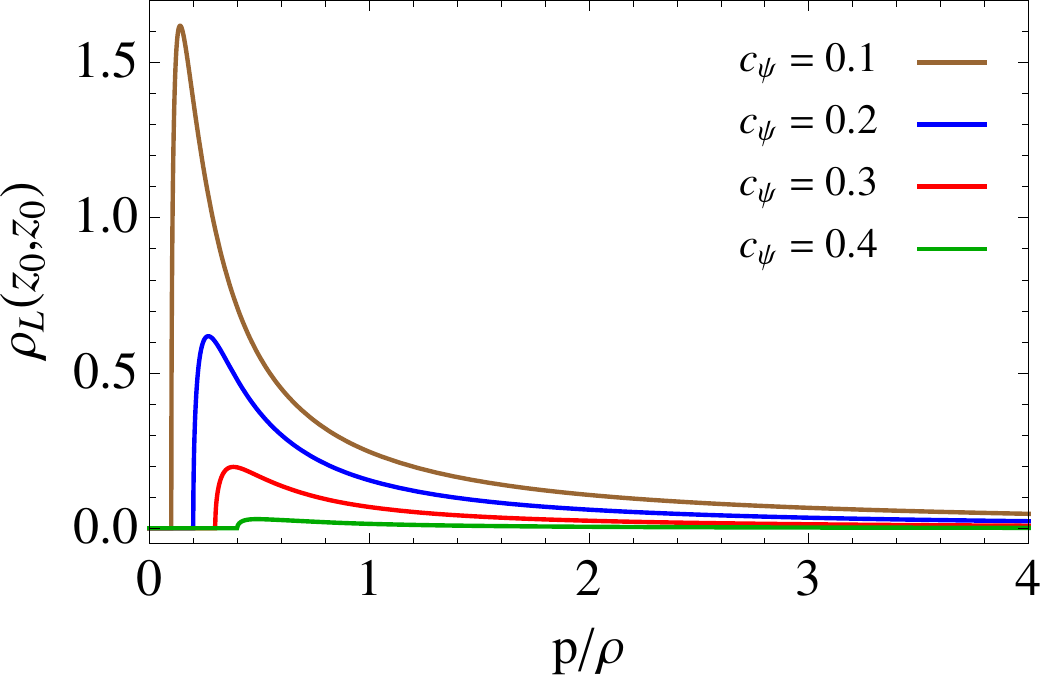} \hspace{1.0cm}
\includegraphics[width=7cm]{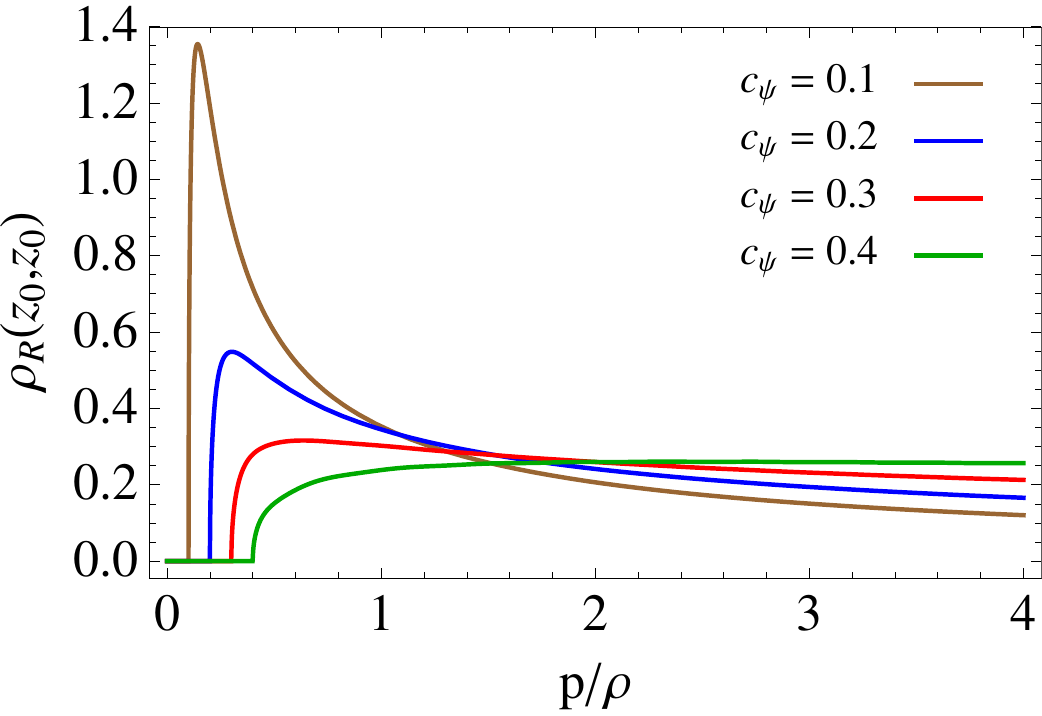} \\
\includegraphics[width=7cm]{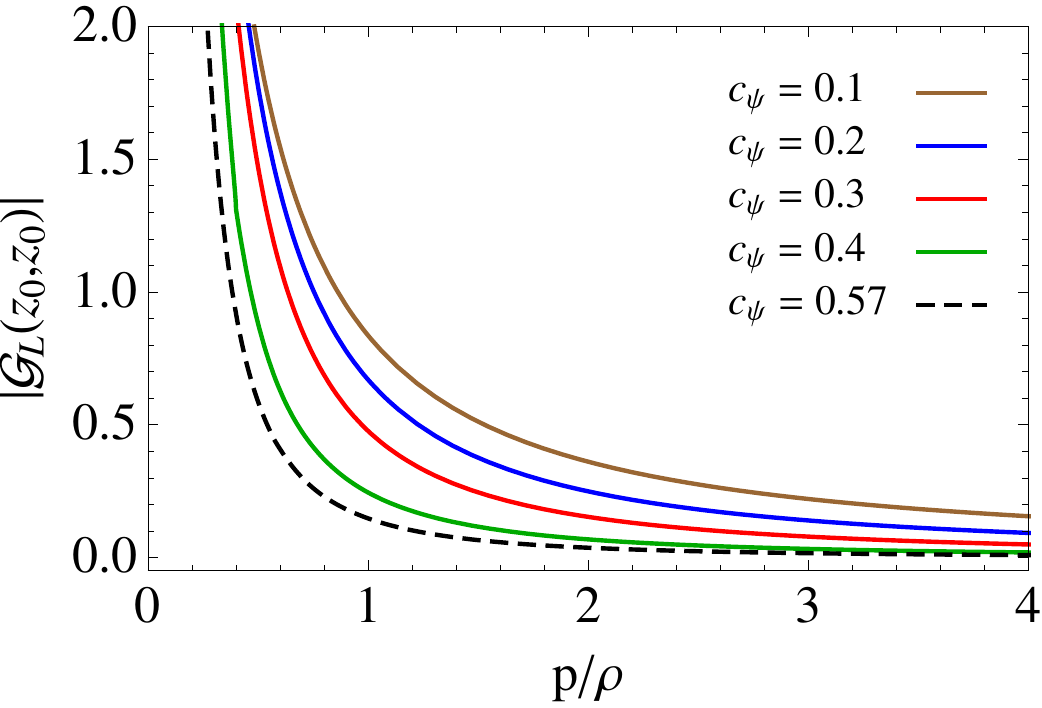} \hspace{1.0cm}
\includegraphics[width=7cm]{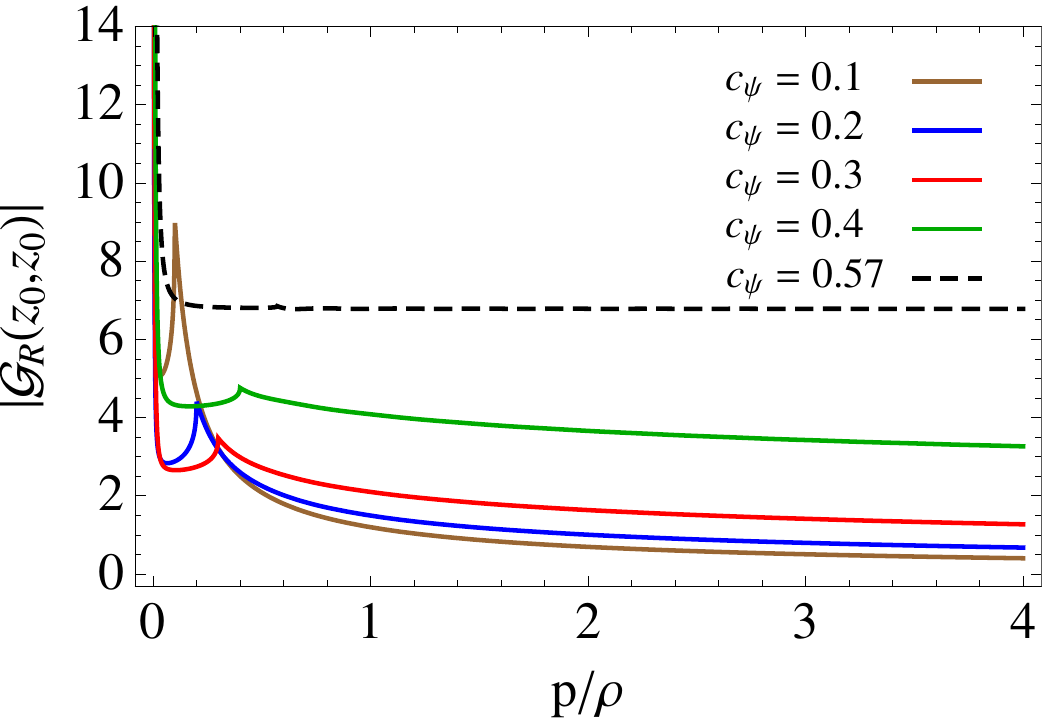}   \\
\caption{\it Upper panels: Spectral density for a continuum left-handed fermion, cf. Eq.~(\ref{eq:sigma_L}) (left panel), and a continuum right-handed fermion, cf. Eq.~(\ref{eq:sigma_R}) (right panel). We have used $c_\psi = 0.1, 0.2, 0.3$ and $0.4$. Lower panels: Absolute value of the Green's functions for a continuum left-handed fermion (left panel), and a continuum right-handed fermion (right panel). We have used $c_\psi = 0.1, 0.2, 0.3, 0.4$ (solid lines) and $0.57$ (dashed lines). All cases use $\epsilon=-1$. The results with $\epsilon=1$ would be the above plots but with $L\leftrightarrow R$. We have used $A_1 = 35$ and $c=1$.
}
\label{fig:spectralfermions2}
\end{figure} 

Similarly to the procedure followed for gauge bosons in
  Sec.~\ref{subsec:gauge_bosons}, we have looked for the appropriate
  rescalings of the Green's functions $G_{L,R}$ which make them scale
  invariant $\mathcal G_{L,R}$. The result is 
  \be
  \mathcal G_{L,R}(z_0,z_0;p)\equiv \rho^{1\pm 2 c_\psi}G_{L,R}(z_0,z_0;p)\,,
  \label{eq:G_scale_invariant}
\ee
where the plus (minus) sign corresponds to $G_L$ ($G_R$). The scale
invariant spectral function $\rho_{L(R)}$ for a left-handed (right-handed) fermion
\be
\rho_{L,R}(z_0,z_0;p)\equiv \rho^{1\pm 2 c_\psi}\sigma_{L,R}(z_0,z_0;p)\,,
\label{eq:rho_scale_invariant}
\ee
is shown in the upper left (right) panel of
Fig.~\ref{fig:spectralfermions2}, for $\epsilon=-1$. Similarly, the
absolute value of the scale invariant Green's function for a
left-handed (right-handed) fermion is displayed in the lower left
(right) panel of Fig.~\ref{fig:spectralfermions2}. The latter
translate into the behavior for the Green's functions $G_{L,R}\sim
p^{-1\mp 2 c_\psi}$ for $p\gg\rho$. 

By using the correlators in the CFT of spinors $\mathcal O_{R,L}$
\begin{align}
\Delta_R(p,c_\psi)&=\frac{\delta^2 S}{\delta \bar\psi^0_L\delta \psi^0_L}=\langle \mathcal O_R\bar{\mathcal O}_R  \rangle\equiv\bar\sigma^\mu p_\mu \bar G_R \,,  \\
\Delta_L(p,c_\psi)&=\frac{\delta^2 S}{\delta \bar\psi^0_R\delta \psi^0_R}=\langle \mathcal O_L\bar{\mathcal O}_L  \rangle\equiv\sigma^\mu p_\mu \bar G_L \,,
\end{align}
their comparison with the propagators in the case of unparticles $\mathcal O_{L,R}$ with dimension~$d$
\be
\Delta_L\propto (-p^2-i\epsilon)^{d-5/2}\sigma^\mu p_\mu\,, \qquad \Delta_R\propto (-p^2-i\epsilon)^{d-5/2}\bar\sigma^\mu p_\mu  \,,
\ee
translates into the dimension for the operators $\mathcal O_{L,R}$: $d_{L,R}=2\mp c_\psi$, in agreement with general results~\cite{Cacciapaglia:2008ns}.

Up to now we have used $\epsilon=-1$ and $c_\psi<1/2$ in Eq.~(\ref{eq:VLR}). It is worth mentioning that a change in the sign of the parameter $\epsilon$, i.e.~considering $\epsilon=1$, is equivalent to the exchange between left-handed and right-handed fermions. This means that the expressions of Eqs.~(\ref{eq:sigma_L}) and (\ref{eq:sigma_R}) for the spectral densities $\sigma_L(p^2)$ and $\sigma_R(p^2)$ would still be valid for $\epsilon=1$, but the results would be as in Fig.~\ref{fig:spectralfermions2}, right panel and left panel, respectively. The same applies for the 4D Green's functions $\mathcal G_L$ and $\mathcal G_R$.

For the case $c_\psi>1/2$ we find that the scale invariant Green's function and spectral density are given by
\begin{align}
  \mathcal G_{L,R}(z_0,z_0;p)&\equiv \rho^{1\pm 2 c_0}G_{L,R}(z_0,z_0;p),\quad c_0=1/2\nonumber\\
  \rho_{L,R}(z_0,z_0;p)&\equiv \rho^{1\pm 2 c_0}\sigma_{L,R}(z_0,z_0;p)\,, \quad c_0=1/2
\end{align}
which coincide with the scaling behaviors in Eqs.~(\ref{eq:G_scale_invariant}) and (\ref{eq:rho_scale_invariant}) with $c_\psi=c_0=1/2$.
We can see this behavior for the Green's functions in the lower panels of Fig.~\ref{fig:spectralfermions2} where we plot the case $c_\psi=0.57$ (dashed lines). 
In the holographic interpretation, the dimension of the CFT operator for $c_\psi>1/2$ is then $d_{L,R}=2\mp c_0$. For the case of $\epsilon=-1$ we are considering here, the 5D fermion has a left-handed massless zero mode with dimension $d_L=3/2$ pointing out toward the existence of an elementary fermion.

\subsection{The graviton}
\label{subsec:graviton}

The graviton is a transverse traceless fluctuation of the metric of the form
\begin{equation}
ds^2 = e^{-2A(y)} (\eta_{\mu\nu} + h_{\mu\nu}(x,y)) dx^\mu dx^\nu - dy^2 \,,
\end{equation}
where $h_\mu^\mu = \partial_\mu h^{\mu\nu} = 0$. We will use the ansatz $h_{\mu\nu}(x,y) =\h(y) h_{\mu\nu}(x) $. The Lagrangian is given by
\be
\mathcal L=-\frac{1}{8 \kappa^2}\int_0^{y_s} dy e^{-2A}\left[\partial_\rho h_{\mu\nu}\partial^\rho h^{\mu\nu}+e^{-2A}h'_{\mu\nu}h^{\prime \mu\nu}
\right] \,,
\ee
from where the EoM can be written as
\be
e^{2A} (e^{-4A}\h^\prime)^\prime+p^2\h(y)=0 \,.
\ee
In conformal coordinates, cf.~Eq.~(\ref{eq:conformal_coord}), and after rescaling the field by ${\h}(z) =e^{3 A(z)/2} \hat\h(z)$, the equation of motion for the fluctuation can be written in the Schr\"odinger like form~\cite{Cabrer:2009we} as
\begin{equation}
-\ddt{\hat\h}(z) + V_{\h}(z) {\hat\h}(z) = p^2 {\hat\h}(z) \,,
\end{equation}
where the potential is given by
\begin{equation}
V_\h(z) = \frac{9}{4} \dt{A}^2(z) - \frac{3}{2} \ddt{A}(z)  \,. \label{eq:Vh}
\end{equation}
Using the EoM of the background, this potential can be expressed in the form
\begin{eqnarray}
V_\h(z) =  \frac{\kappa^2}{48} e^{-2A(z)} \left( 5 \kappa^2 W^2[\phi(z)] - 6 \left(W^\prime[\phi(z)]\right)^2  \right) \,.
\label{eq:potgraviton}
\end{eqnarray}
This potential has a mass gap, and its value is given by
\begin{equation}
V_\h(z) \stackrel[z \to \infty]{\longrightarrow}{} \frac{9}{4}\rho^2 \,.
\end{equation}

The spectral density and Green's function are obtained by again using holographic methods. In momentum space the graviton field is decomposed as
\be
h_{\mu\nu}(p,z)=\h(p,z) h_{\mu\nu}^{(4)}(p) \,,
\ee
where $h_{\mu\nu}^{(4)}(x)$ is a 4D graviton field whose components satisfy the Klein-Gordon equation
\be
(\Box+p^2)h_{\mu\nu}^{(4)}(x)=0 \,,
\ee
as well as conditions $\partial^\mu h^{(4)}_{\mu\nu}=0$ and $h^{\mu(4)}_{\mu}=0$, which eliminate five out of the ten components of the symmetric tensor $h^{(4)}_{\mu\nu}$.
By replacing the bulk EoMs we obtain the holographic Lagrangian
 \be
 \mathcal L_{hol}=\frac{1}{32\kappa^2} \h(p,z_0)\dt{\h}(p,z_0) h_{\mu\nu}^{(4)}(p){h}_{\mu\nu}^{(4)}(p) \,.
 \ee
After fixing the boundary conditions at the UV brane in terms of the source field $h_{\mu\nu}^0(p)$, 
 \be
 h_{\mu\nu}(p,z_0)=h_{\mu\nu}^0(p)\,,
 \ee
 coupled to the CFT operator $\mathcal O_{\mu\nu}$, the holographic Lagrangian reads as
\be
 \mathcal L_{hol}=\frac{1}{32\kappa^2}\frac{\dt{\h}(p,z_0)}{\h(p,z_0)}h_{\mu\nu}^0(p)h_{\mu\nu}^0(p) \,,
\ee
from where the Green's function and spectral density are given by
\be
 G_\h (z_0,z_0;p)=\frac{\h(p,z_0)}{\dt{\h}(p,z_0)} \,,\qquad \sigma_\h(z_0,z_0;p)=\frac{1}{\pi}\textrm{Im}\left[ \frac{\h(p,z_0)}{\dt{\h}(p,z_0)}  \right] \,.
\ee
 
\begin{figure}[htb]
\centering
\includegraphics[width=7cm]{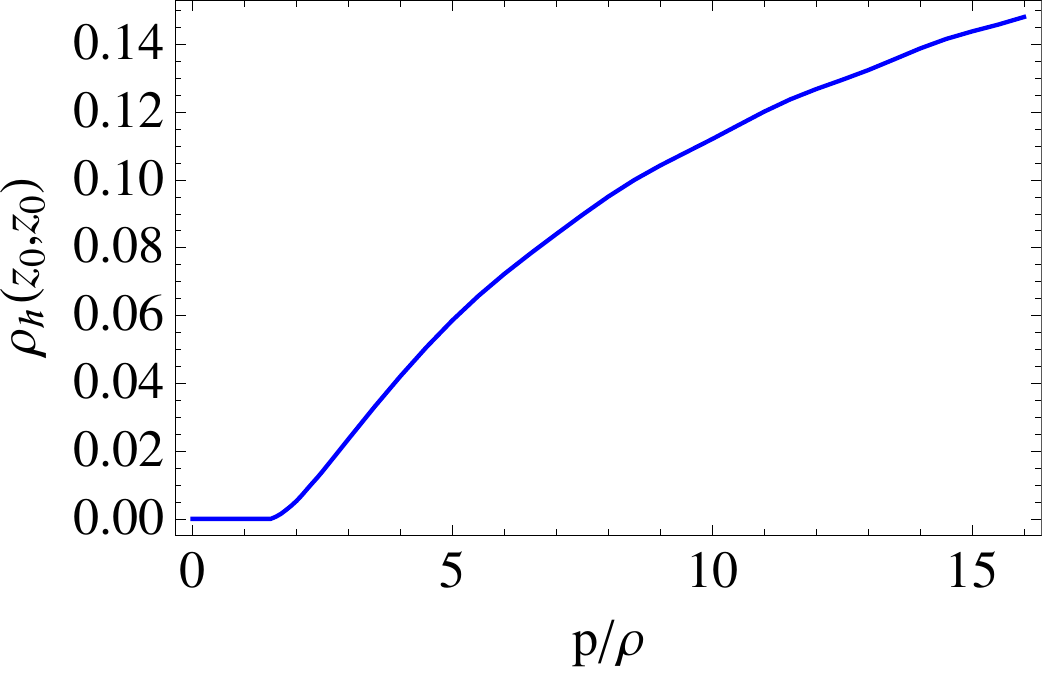} \hspace{0.5cm}
\includegraphics[width=7cm]{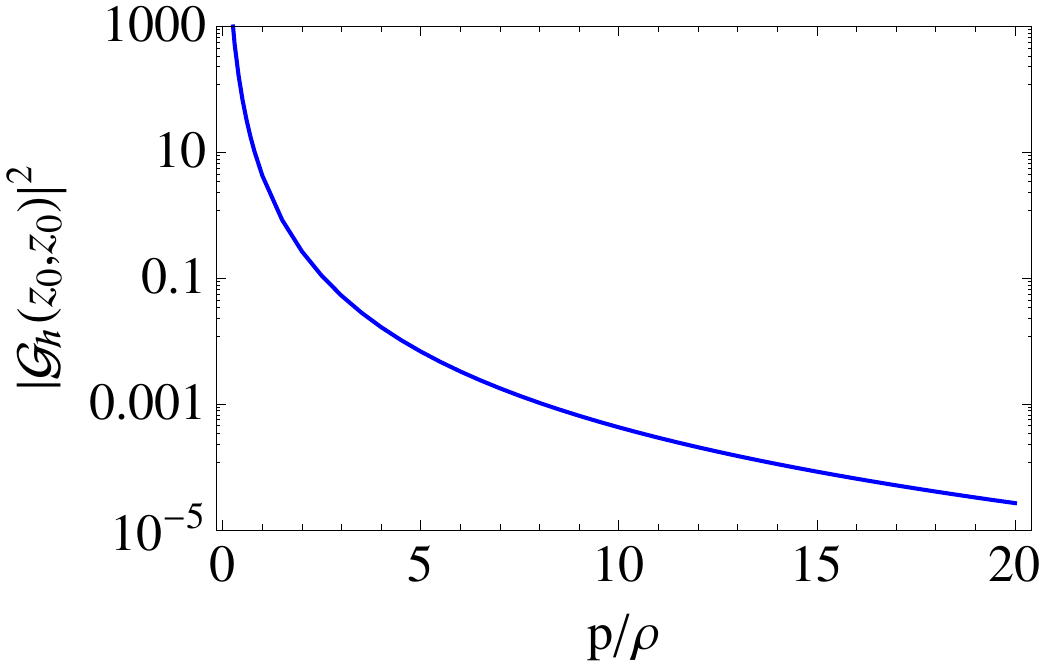}
\caption{\it Left panel: Spectral density $\rho_\h(z_0,z_0;p)$ for a continuum graviton. Right panel: Absolute value squared of the scale invariant Green's function, $\mathcal G_\h(z_0,z_0;p)$, for the graviton. We have used $A_1 = 35$ and $c=1$.}
\label{fig:spectral_gravitons}
\end{figure} 
A plot of the spectral function $\rho_\h=k \sigma_\h$ is shown in the left panel of Fig.~\ref{fig:spectral_gravitons}. The scale invariant Green's function 
\be
\mathcal G_\h(z_0,z_0;p)\equiv (\rho^2/k) G_\h(z_0,z_0;p) 
\ee
is shown in the right panel of Fig.~\ref{fig:spectral_gravitons}. The required rescaling, translates into the behavior $\mathcal G_\h\sim \tau_\h/p^2$ for $p\gg\rho$ where 
\be
\tau_\h  \simeq  \frac{4 \cW^2(k/\rho)}{1+2 \cW(k/\rho) [\cW(k/\rho)-1]}  =  \frac{2 M^3}{k M_P^2} \,,
\ee
which is $\tau_\h \approx 2.1$ for $A_1 = 35$ and $c=1$. In the IR region the Green's function shows an isolated massless pole corresponding to the graviton mode, as can be seen from the right panel of Fig.~\ref{fig:spectral_gravitons}.

\subsection{The radion}

The radion field $\xi(x,y)$ is defined as the metric perturbation
\be
\phi(x,y)=\phi(y)+\delta\phi \quad \textrm{and}\quad ds^2=-N^2dy^2+g_{\mu\nu}(dx^\mu+N^\mu dy)(dx^\nu+N^\nu dy) \,,
\ee
with $N=1+\delta N$, $N_\mu=\partial_\mu\psi$ and $g_{\mu\nu}=e^{-2A-2\,\xi}\,\eta_{\mu\nu}$. Here we will consider the unitary gauge $\delta\phi=0$ and follow the approach of Ref.~\cite{Megias:2014iwa}~\footnote{For the choice of a different gauge where $N_\mu=0$ and $\delta\phi\neq 0$ see Ref.~\cite{Csaki:2000zn}.}. After using the equations of motion, $\delta N$ and $N_\mu$ can be obtained in terms of $\xi(x,y)$ and its derivatives. Then using the background solution one can cast the action as
\be
S=\frac{1}{2\kappa^2}\int d^5x  e^{-4A}\beta^2\left[ e^{2A}(\partial_\mu \xi)^2-(\xi^\prime)^2  \right] \,,
\ee
where $\beta=-\kappa \phi^\prime/A^\prime$, leading to the bulk EoM
\begin{equation}
e^{2A}\frac{1}{\beta^2}  \left[e^{-4A}\beta^2 \xi^\prime  \right]^\prime=\Box \xi \,.%+ e^{-2A}  C \xi \delta(y-y_0) \,,
\end{equation}
In conformal coordinates and after redefining $\xi=e^{(3/2)A} \hat\xi/\beta$, one can cast the EoM in a Schr\"odinger-like form, as
\be
-\ddt{ \hat\xi}(z) + V_\xi(z) \hat\xi(z) = p^2 \hat\xi(z) \,,
\ee
where the potential 
\be
V_\xi=\beta^2G\left(G-\frac{\dt\beta}{\beta^2}\right)-\beta \dt G,\quad \textrm{with}\quad G=-\frac{\dt\beta}{\beta^2}+\frac{3}{2}\frac{\dt A}{\beta}
\ee
goes to $9\rho^2/4$ in the limit $z\to \infty$.

The spectral density can again be obtained by using holographic methods. After replacing the EoM into the action one obtains the holographic Lagrangian in momentum space as 
\be
\mathcal L_{hol} = \frac{1}{2\kappa^2} e^{-3A(z_0)} \left[ \beta^2(z_0)f(p,z_0)\dt f(p,z_0) -8 \bar\lambda_0(v_0)  f^2(p,z_0)\right] \mathcal R^{(4)}\mathcal R^{(4)} \,,
\ee
where $\bar\lambda_0\equiv 2\kappa^2\lambda_0$ and we have decomposed $\xi(p,z)=f(p,z)\mathcal R^{(4)}(p)$ with the 4D scalar field $\mathcal R^{(4)}$ satisfying the free field equation.
After fixing the UV condition
\be
\xi(p,z_0)\equiv \mathcal R^0(p) \,,
\ee
one can write 
\be
\mathcal L_{hol}=  \frac{1}{2\kappa^2} \left[ \frac{3}{(1 + ky_s)^2}  \frac{\dt f(p,z_0)}{f(p,z_0)} - 8 \bar\lambda_0(v_0) \right] \mathcal R^0(p) \mathcal R^0(p) \,.
\ee
Then the Green's function for the radion reads as
\be
G_\xi(z_0,z_0;p)= \left( \frac{3}{(1 + ky_s)^2}  \frac{\dt f(p,z_0)}{f(p,z_0)}  - 8 \bar\lambda_0(v_0) \right)^{-1} \,, \label{eq:GreenFunction_radion}
\ee
 and the spectral density is given by
\be
\sigma_\xi(z_0,z_0;p) = \frac{1}{\pi} \textrm{Im} \, G_\xi(z_0,z_0;p) \,. \label{eq:spectral_radion}
\ee
\begin{figure}[htb]
\centering
\includegraphics[width=7cm]{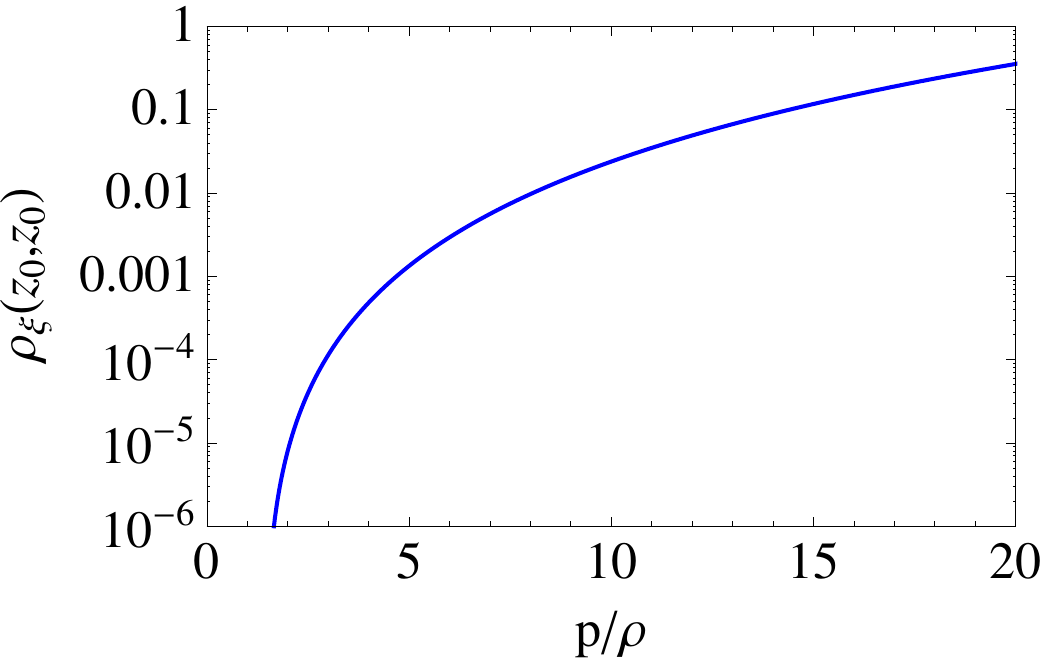}   \hspace{0.5cm}
\includegraphics[width=7cm]{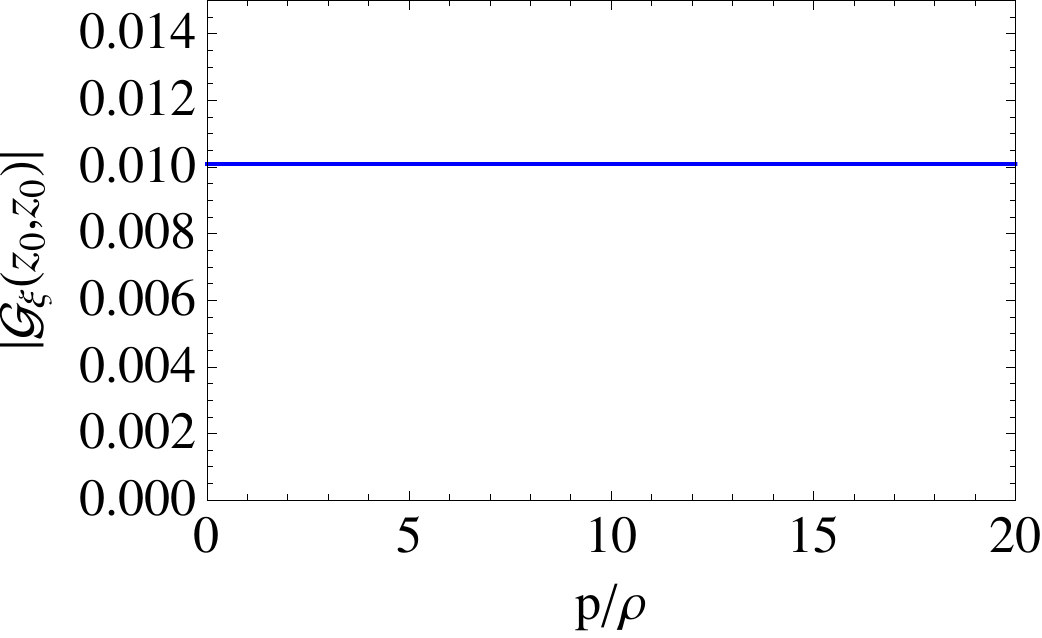} 
\caption{\it
Left panel: Scale invariant spectral density for a continuum radion, as given by Eq.~(\ref{eq:spectral_radion_si}). Right panel: Scale invariant Green's function for a continuum radion, as given by Eq.~(\ref{eq:G_xi}). We have used $A_1=35$ and $c=1$. 
}
\label{fig:spectral_radion}
\end{figure} 
Note that the UV boundary conditions in the background demand (cf. Eq.~(\ref{eq:BCWprime})) 
\begin{equation}
\bar\lambda_0(v_0) =2\kappa^2 W(\phi_0) = 12 k \left(1 + \frac{1}{ky_s} \right) \,.
\end{equation}
The results for the scale independent spectral density
\be
\rho_\xi(z_0,z_0;p)\equiv k \left( \rho/k \right)^{-4} \sigma_\xi(z_0,z_0;p) \,, \label{eq:spectral_radion_si}
\ee
and Green's function
\be
\mathcal G_\xi(z_0,z_0;p) \equiv  k G_\xi(z_0,z_0;p)\,,  \label{eq:G_xi}
\ee
are plotted in the left and right panels, respectively, of Fig.~\ref{fig:spectral_radion}. The Green's function has a constant behavior at momenta $p \sim {\cal O}(\rho)$, as the term $\lambda_0(v_0)$ is by far the dominant contribution at these scales. More in detail, the behavior of the Green's function at low momenta is
\begin{equation}
\mathcal G_\xi^{-1}(z_0,z_0;p)  \simeq  \frac{1}{\tau_\xi} \mathcal W^{-2}(k/\rho)  (\rho/k)^{2}  (p/\rho)^2 - 8 k^{-1} \bar\lambda_0(v_0) \,,  \label{eq:Gradionp}
\end{equation} 
where 
\be
\tau_\xi = \frac{2}{3} \frac{(1 + ky_s) e^{2k y_s}}{( 1 + ky_s +(-1+ky_s) e^{2k y_s} )} \simeq \frac{2}{3} \frac{{\mathcal W}(k/\rho) + 1}{{\mathcal W}(k/\rho) - 1} \,,
\ee
which is $\tau_\xi \simeq 0.71$  for $A_1 = 35$ and $c=1$. In Eq.~(\ref{eq:Gradionp}) we have neglected an imaginary contribution, which is typically small. Only when $p$ is large enough, the Green's function has a non-constant behavior and in particular it goes to zero when $p\to\infty$.

\subsection{The Higgs boson}

The action for the physical Higgs boson $H(x,y)\equiv h(y)+\widehat H(x,y)$ is given by Eq.~(\ref{eq:S5Higgs}), from where the bulk EoM, UV boundary and IR jump conditions for the excitation $\widehat H(x,y)$ are
\begin{align}
&  \widehat H^{\prime\prime}(y) - 4 A^\prime(y) \widehat H^\prime(y) - \frac{\partial^2 V(h)}{\partial h^2}\widehat H+ p^2 e^{2A}\widehat H(y) = 0 \,, \nonumber\\
& \widehat H^\prime(0)=\left.\frac{1}{2} \frac{\partial^2 \lambda^0(h)}{\partial h^2} \widehat H\right|_{y=0}, \qquad
\frac{\Delta\widehat  H^\prime(y_1)}{\widehat H(y_1)} =-\left. \frac{\partial^2 \lambda^1(h)}{\partial h^2}\right|_{y=y_1}\,,
\end{align}
where we have already replaced the squared mass eigenvalue $m^2$ by $p^2$, and used the equation of motion (\ref{eq:h}), boundary condition (\ref{eq:UVBCh}) and jump condition (\ref{eq:IRBCh}), for the background field $h(y)$. The bulk equation can be expressed in a Schr\"odinger like form, leading to the result
\begin{equation}
- \ddt{\widehat H}(z) + V_H(z) \widehat H(z) = p^2 \widehat H(z) \,, \ \textrm{with} \   V_H(z) =  \frac{9}{4} \dt{A}^2(z) - \frac{3}{2} \ddt{A}(z) + e^{-2A(z)} \frac{\partial^2 V(h)}{\partial h^2}  \,. \label{eq:EoMHiggsFluctuations}
\end{equation}
In this equation we have rescaled the field as $\widehat H(z) \to e^{3A(z)/2} \widehat H(z)$. Eq.~(\ref{eq:EoMHiggsFluctuations}) yields a mass gap of $(3\rho/2)^2$.

 After Fourier transforming the configuration space $x^\mu$ into momentum space $p^\mu$, and going to conformal coordinates $z$, we can decompose the field as
\be
 H(p,z)=\mathcal H(p,z) \mathcal H^{(4)}(p),\quad \textrm{where}\quad (\Box+p^2)\mathcal H^{(4)}(x)=0 \ .
\ee
Replacing now the EoM, we obtain the holographic Lagrangian as
\be
\mathcal L_{hol}=\frac{1}{2}\left[ \mathcal H(p,z_0)\dt{\mathcal H}(p,z_0)-2 M_0 \mathcal H^2(p,z_0)
\right] \mathcal H^{(4)}(p)\mathcal H^{(4)}(p) \,.
\ee 
We now fix the boundary conditions at the UV brane in terms of the source field $H^0(p)$ as
\be
H(p,z_0)=\mathcal H(p,z_0) \mathcal H^{(4)}(p)\equiv H^0(p) \,,
\ee
so that the holographic Lagrangian is written as
\be
\mathcal L_{hol}=\frac{1}{2}\left[ \frac{\dt{\mathcal H}(p,z_0)}{\mathcal H(p,z_0)}-2 M_0\right] H^0(p) H^0(p) \,.
\ee
Then, finally the Green's function and spectral density can be written as
\begin{eqnarray}
G_H(z_0,z_0;p) &=& \left( \frac{\dt{\mathcal H}(p,z_0)}{\mathcal H(p,z_0)}-2 M_0\right)^{-1}\,, \label{eq:GH} \\
\sigma_H(z_0,z_0;p) &=& \frac{1}{\pi}\textrm{Im}\left( \frac{\dt{\mathcal H}(p,z_0)}{\mathcal H(p,z_0)}-2 M_0\right)^{-1}\,.
\end{eqnarray}
\begin{figure}[htb]
\centering
\includegraphics[width=6.9cm]{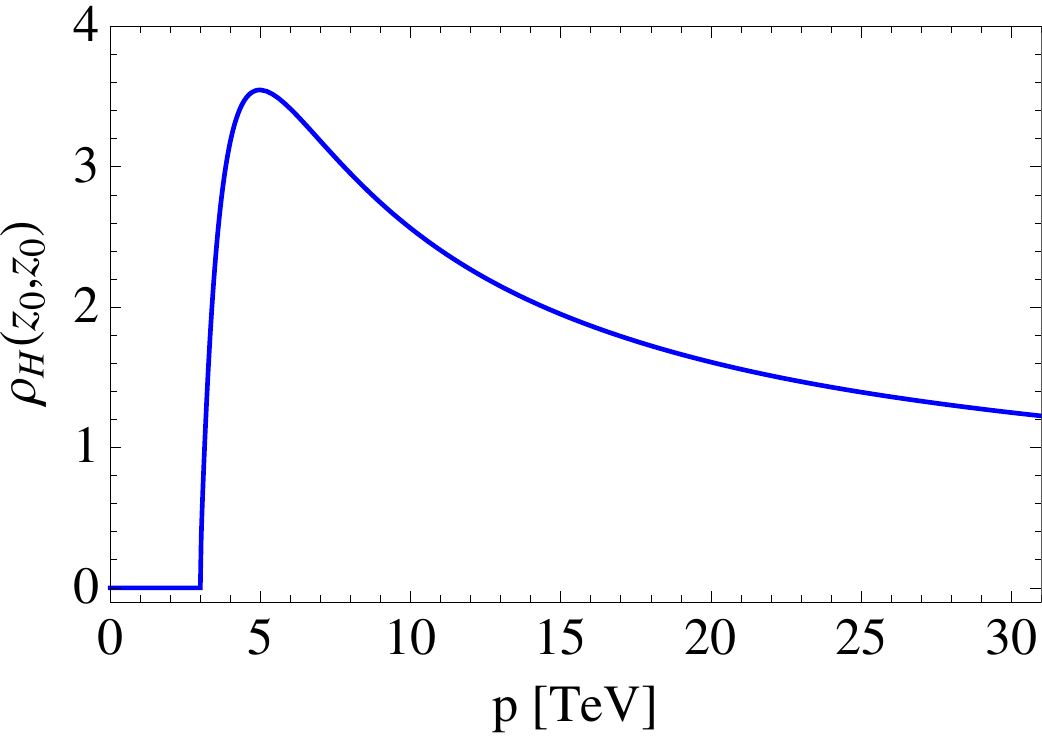} \hspace{1.0cm}
\includegraphics[width=7.4cm]{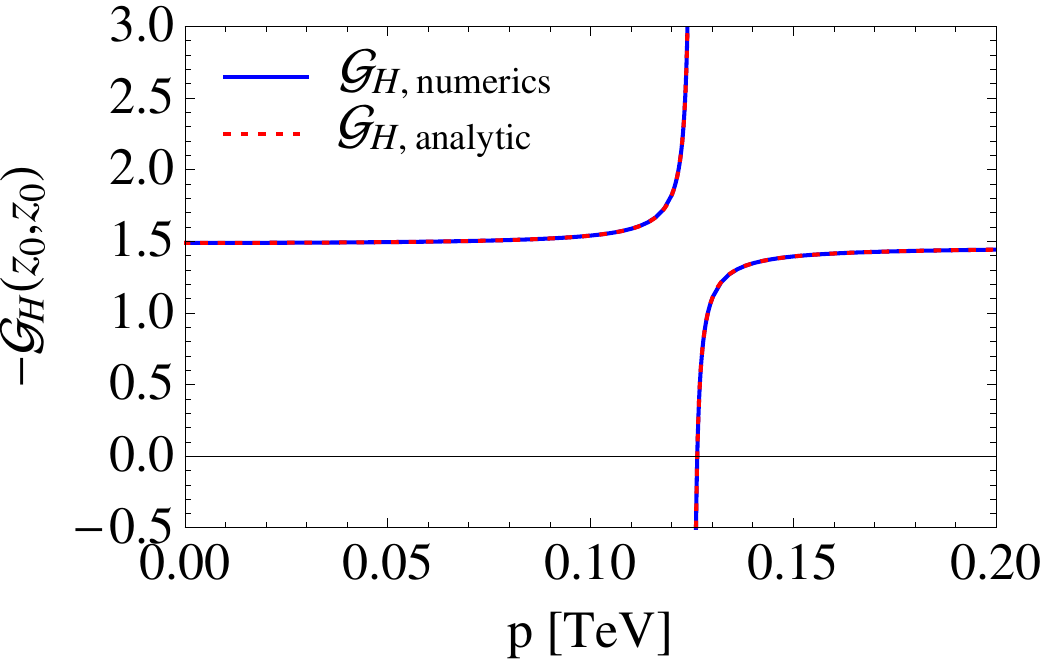}
\caption{\it Left panel: Spectral density $\rho_H(z_0,z_0;p)$ for a continuum Higgs boson. Right panel: Green's function, $\mathcal G_H(z_0,z_0;p)$, for the Higgs. We display in solid (blue) line the numerical computation using Eq.~(\ref{eq:GH}), and in dotted (red) line the analytical result from Eq.~(\ref{eq:GHanalytic}). We have used $A_1 = 35$, $c=1$, $ \rho = 2$ TeV, $a=2.4$ and $M_0 = 0.5 \cdot ka$. The pole of the Green's function is located at $p = 0.125$ TeV.
}
\label{fig:GreenFunction_Higgs}
\end{figure} 
The scale invariant spectral 
\be
\rho_H(z_0,z_0;p)\equiv k(\rho/k)^{2-a}\sigma_H(z_0,z_0;p)
\ee
and Green's functions 
\be
\mathcal G_H(z_0,z_0;p)\equiv k G_H(z_0,z_0;p)
\ee
are plotted in the left and right panels of Fig.~\ref{fig:GreenFunction_Higgs}. The latter in particular implies for the Green's function a constant behavior away from the pole position. In particular the value of the Green's functions for $p > m_H$ is
\begin{equation}
\mathcal G_H^{(+)}(z_0,z_0)^{-1}  =  - \frac{2M_0}{k} + a - \frac{1}{{\bar F}(y_1)}  \,, \label{eq:G_pinf}
\end{equation}
and for  $p < m_H$ is
\begin{equation}
\mathcal G_H^{(-)}(z_0,z_0)^{-1}  \simeq \mathcal G_H^{(+)}(z_0,z_0)^{-1} + \frac{{\bar F}^\prime(y_1)}{2M_1 {\bar F}(y_1)^2}  + \cdots   \,, \label{eq:G_minf}
\end{equation}
with ${\bar F}(y) \equiv F(y) - F(0)$, where $F(y)$ is defined in Eq.~(\ref{eq:F}). After performing an expansion for $k y_s \gg 1$, one finds
\begin{equation}
\mathcal G_H^{(+)}(z_0,z_0)^{-1}  \simeq  -\frac{2M_0}{k}  + (4-a) + \frac{4}{k y_s} + \frac{2}{(a-2) (k y_s)^2}  + \cdots  \,,
\end{equation}
and
\begin{eqnarray}
\mathcal G_H^{(-)}(z_0,z_0)^{-1}  &\simeq&   \mathcal G_H^{(+)}(z_0,z_0)^{-1} \label{eq:G_minf2} \\
&&+ 16 (a-1)^3 (a-2)^2 (k y_s)^4  \frac{\rho^2}{m_H^2} e^{-2(a-2)ky_s} \left( 1 - \frac{4}{(a-2) k y_s} + \cdots \right)   \,.  \nonumber  
\end{eqnarray}
Moreover,  for $p\simeq m_H$, the Green's function shows a pole behavior as $$\mathcal G_H\sim m_H^2/(p^2-m_H^2)$$ which corresponds to the presence of the isolated resonance corresponding to the presence of the SM Higgs. This pole is shown in the right panel of Fig.~\ref{fig:GreenFunction_Higgs} where we have fixed $\rho=2$ TeV, and we have considered $ky_1 =k y_s -\cW(c)$.
In fact an excellent agreement with the full numerical behavior of the Green's function is given by
\be
\mathcal G_H(z_0,z_0;p) = c_1 + c_2 \frac{m_H^2}{p^2-m_H^2} \,,  \label{eq:GHanalytic}
\ee
where the coefficient $c_1 =  \mathcal G_H(z_0,z_0;p\to\infty)$ is determined analytically from Eq.~(\ref{eq:G_pinf}), and 
\begin{eqnarray}
c_2 &=&  \mathcal G_H(z_0,z_0;p \to \infty) -  \mathcal G_H(z_0,z_0;p \to 0)  \\
&\simeq& \frac{{\bar F}^\prime(y_1)}{2 M_1 {\bar F}(y_1)^2} \mathcal G_H^{(+)}(z_0,z_0)^{2} \simeq  \frac{\rho^2}{m_H^2} \frac{16 (a-1)^3 (a-2)^2 (k y_s)^4 e^{-2(a-2) k y_s}}{\left(-\frac{2M_0}{k} + 4-a\right)^2}   + \cdots  \nonumber 
\end{eqnarray}
is a relatively small difference between the UV and IR limits of the Green's function. In the second and third equalities of this equation we have used Eqs.~(\ref{eq:G_minf}) and (\ref{eq:G_minf2}) respectively. The values of these coefficients for the plot of the right panel of Fig.~\ref{fig:GreenFunction_Higgs} are $c_1=-1.46$ and $c_2= 2.88\times 10^{-2}$, and the corresponding result is shown as a dotted line in that figure.

\section{Phenomenological aspects}
\label{sec:Greens_functions}

The main experimental signature for detecting heavy new physics is its production, and subsequent decay, in colliders, and in particular in the present LHC at CERN. When the new physics consists in heavy Breit-Wigner resonances, their presence is detected by bumps in the invariant mass of the final states, corresponding to the mass of the exchanged particle. However, when the new physics consists in a continuum of states beyond a mass gap~$m_g$, as in the model we are considering in this paper, its presence should be associated, not with a bump but with an excess, with respect to the Standard Model prediction, in the measured cross section. The larger the mass gap, the higher energy should one produce to detect the excess in the predicted cross sections. In this way the continuum states with the least mass gap are the most easily produced. As we have shown in the previous section, the different mass gaps for the different fields are those summarized in Tab.~\ref{table}.
\begin{table}[ht]
\centering
\begin{tabular}{||c||c|c|c|c|c||}
\hline\hline
Field & Gauge boson & Fermion $f$& Graviton & Radion & Higgs \\
\hline
$m_g$ & $\rho/2$ & $|c_f|\rho$ & $3\rho/2$ & $3\rho/2$& $3\rho/2$\\
\hline\hline
\end{tabular}
\caption{\it Values of the mass gap for different fields where $\rho\equiv e^{-ky_s}/y_s$.}
\label{table}
\end{table}
From there we can see that, as for light fermions $c_f>1/2$ and so their mass gap is $m_g>\rho/2$, the simplest case for producing the continuum of KK modes are gauge bosons and, in particular the strongest coupled KK modes, the KK gluons, for which we will concentrate ourselves in this section. In Drell-Yan (DY) processes the continuum of KK gluons is produced by pairs of light fermions (valence quarks in the proton), which we can assume to be localized on the UV brane at $z=z_0$. Subsequently, the continuum will decay into a pair of fermions. The more localized toward the IR brane the fermions are, the more strongly they are coupled to KK modes, and the more copiously they are produced. To study the different processes we must compute the gluon Green's function propagating between two arbitrary points in the bulk $G_A(z,z';p)$. 

Up to now we have presented the computation of Green's functions $G_A(z_0,z_0;p)$, and spectral densities, by using the holographic method. There is an alternative procedure to compute these quantities, by obtaining the UV-brane-to-UV-brane Green's functions $G_A(p)$ from the general Green's function $G_A(z,z^\prime;p)$ in the form
\begin{equation}
G_A(z_0,z_0;p) = \lim_{z,z^\prime \to z_0} G_A(z,z^\prime; p) \,,  \label{eq:limGp2}
\end{equation}
where $G_A(z,z^\prime;p)$ is the ``IR regular'' solution to the inhomogeneous bulk EoM. Moreover the general formalism will allow us to compute the case of arbitrary, $z$ and $z'$, and in particular the interesting case where the Green's function propagates from the UV to the IR brane, i.e.~when $z=z_0$ and $z'=z_1$.

In the case of gauge bosons, the general equation writes as 
\begin{equation}
e^{-A} p^2 G_A(z,z';p) + \frac{d}{dz}\left[ e^{-A} \dt{G}_A(z,z';p) \right] = \delta(z-z^\prime) \,, \label{eq:GAz}
\end{equation}
where the dot indicates derivative with respect to $z$,
which is the inhomogeneous version of Eq.~(\ref{eq:fAy})~\footnote{One could equivalently work in proper coordinates. Eq.~(\ref{eq:GAz}) would write as
$$
p^2 G_A(y,y';p) + \frac{d}{dy}\left[ e^{-2A} G_A^\prime(y,y';p) \right] = \delta(y-y^\prime) \,,
$$
where the prime indicates derivative with respect to the variable $y$, and we have taken into account that $\delta(y-y^\prime) = e^A \delta(z-z^\prime)$.
}. This equation includes the generalization to the continuous spectrum of the Green's function, which for the case of a discrete spectrum with mass $m_n$ for the $n$-th mode $f_A^n$, is well-known to be given by
\be
G_A(z,z';p)= \sum_n \frac{f_A^n(z)f_A^n(z^\prime)}{p^2-m_n^2} \,,
\ee
where $\{f_A^n(z)\}$ is a basis of orthonormal modes. 

The Green's function is subject to boundary and matching conditions. In particular, after fixing the value of $z'$, we divide the $z$ space into the following domains: $z_0 \le z \le z^\prime$, $z^\prime \le z \le z_1$ and $z_1 \le z$, and consider the following conditions for the different values of $z$:
\begin{align}
\begin{split}
& \dt{G}_A(z_0) = 0 \,, \qquad \Delta G_A(z^\prime) = 0 \,, \qquad \Delta \dt{G}_A(z^\prime) = e^{A(z^\prime)} \,, \\
& \Delta G_A(z_1) = 0 \,, \qquad \Delta \dt G_A(z_1)  = 0 \,,  \label{eq:GA_bc}
\end{split}
\end{align}
where $\Delta f(z) \equiv  \lim_{\epsilon \to 0} \left( f(z+\epsilon) - f(z-\epsilon) \right)$. In addition, we should impose regularity in the IR as explained below Eq.~(\ref{eq:fA_IR}), i.e.~we consider $c_+ = 0$. The jump in the derivative of the Green's function at $z=z^\prime$ follows after integrating Eq.~(\ref{eq:GAz}) in the interval $[z^\prime - \epsilon, z^\prime + \epsilon]$. Notice that Eq.~(\ref{eq:GAz}) is a second order differential equation, so that there appear two integration constants in each of the three domains mentioned above, leading to a total of six integration constants. These are fixed after considering the five conditions of Eq.~(\ref{eq:GA_bc}) plus one condition of IR regularity. Finally, once the Green's function $G_A(z,z';p^2)$ is computed, the spectral density for gauge bosons is obtained as
\begin{equation}
\sigma_A(z,z^\prime;p)=\frac{1}{\pi}\, \textrm{Im} \, G_A(z,z^\prime;p) \,.
\end{equation}
\begin{figure}[htb]
\centering
\includegraphics[width=7cm]{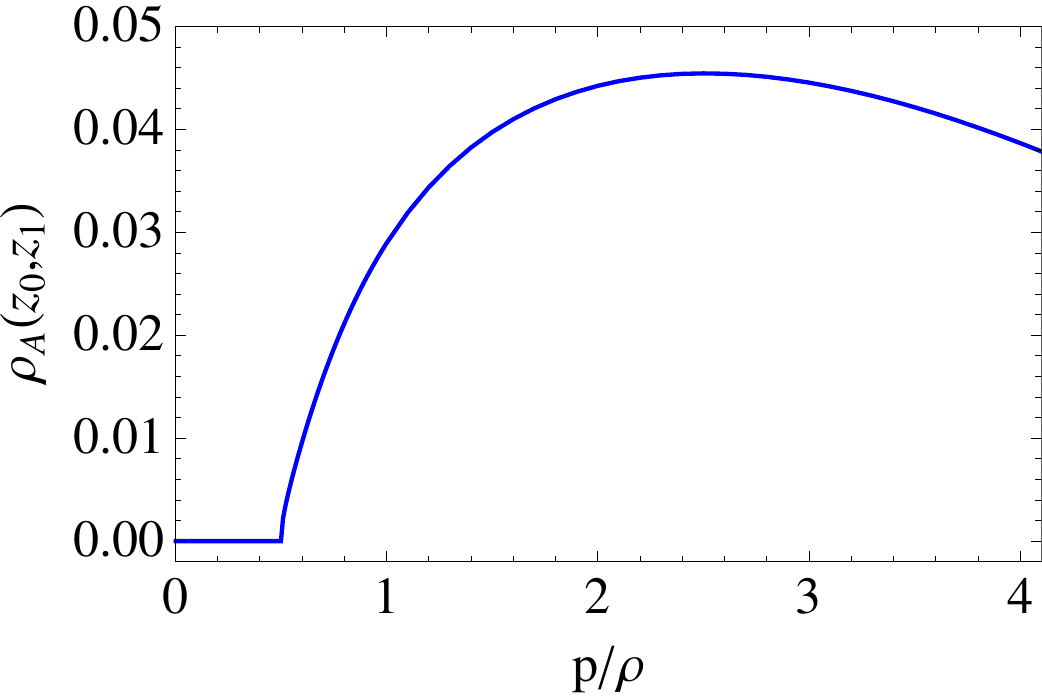} \hspace{0.5cm}
\includegraphics[width=7cm]{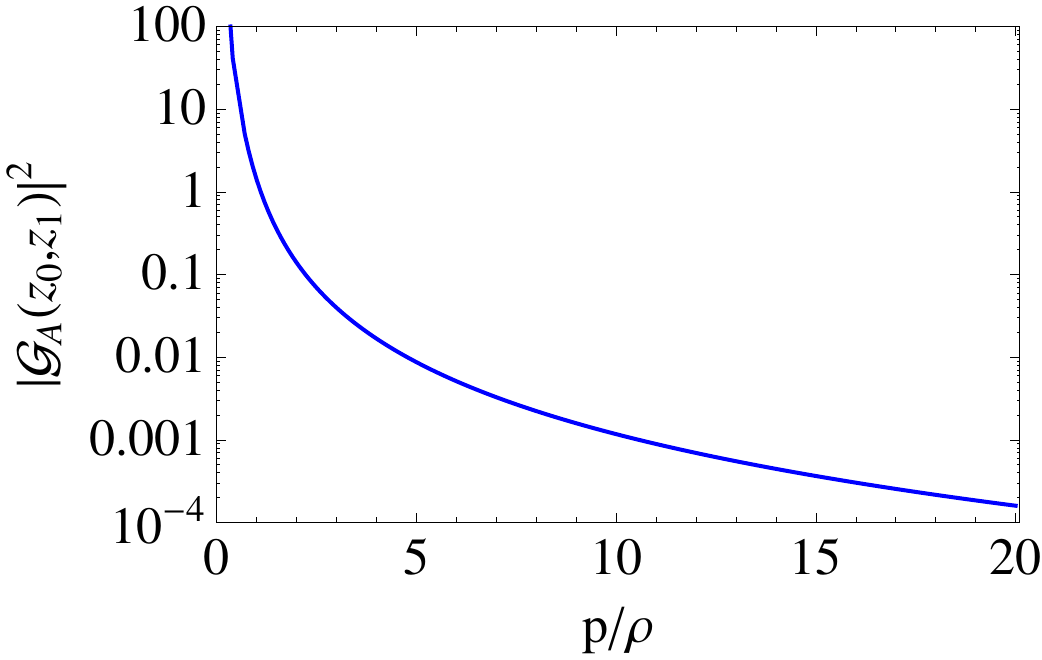} \hspace{0.5cm}
\caption{\it Scale invariant spectral density (left panel) and squared absolute value of the Green's function (right panel) for gauge bosons, for the case $z=z_0$ and $z'=z_1$. We have used $A_1 = 35$ and $c=1$.
}
\label{fig:spectral_gaugebosony1}
\end{figure} 
For the case $z=z^\prime=z_0$ the value of $G_A(p)$ is computed as in Eq.~(\ref{eq:limGp2}), and the result agrees with the plot in the right panel of Fig.~\ref{fig:spectral_gaugebosons}. For the case $z=z_0$ and $z^\prime=z_1$ the scale invariant spectral function
\be
\rho_A(z_0,z_1;p)\equiv\mathcal W(k/\rho)(\rho^2/k) \sigma_A(z_0,z_1;p)
\ee
and Green's function 
\be
\mathcal G_A(z_0,z_1;p)\equiv \mathcal W(k/\rho)(\rho^2/k)G_A(z_0,z_1;p)  
\ee
are given in the left and right panels of Fig.~\ref{fig:spectral_gaugebosony1}, respectively.

In particular the Green's function $\mathcal G_A(z_0,z_1;p^2)$ can be used to compute the contribution of the gauge continuum to the physical partonic process with partonic energy $\sqrt{\hat s}$: $\sigma(q\bar q\to g^*\to Q\bar Q)=\sigma_{SM}(q\bar q\to g^{(0)}\to Q\bar Q)|(\hat s/\rho^2) \, \mathcal G_A(z_0,z_1;\hat s)|^2$, where $g^*$ is the contribution from the gluon continuum, $g^{(0)}$ is the SM gluon and we have made use of the relation between the 5D and 4D gauge couplings $g_5=g_4\sqrt{y_s}$. We are assuming that $q$ is a proton valence quark, living on the UV brane, and $Q$ is either a heavy quark living on the IR brane (as e.g.~$t_R$)
\begin{figure}[htb]
\centering
\includegraphics[width=7.5cm]{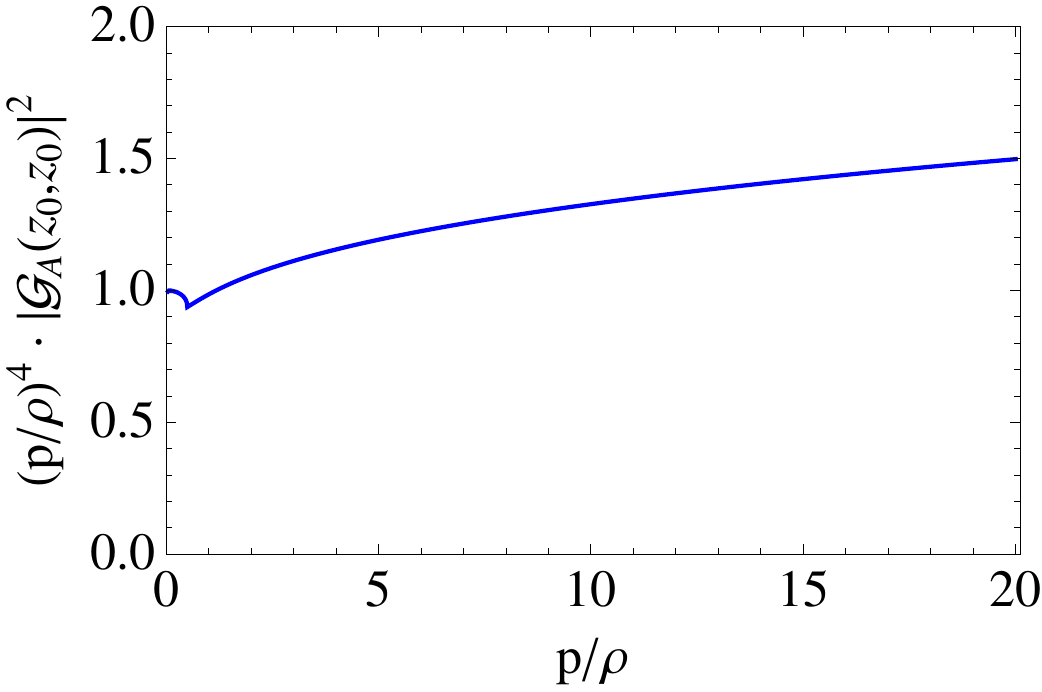} \hspace{0.15cm}
\includegraphics[width=7.5cm]{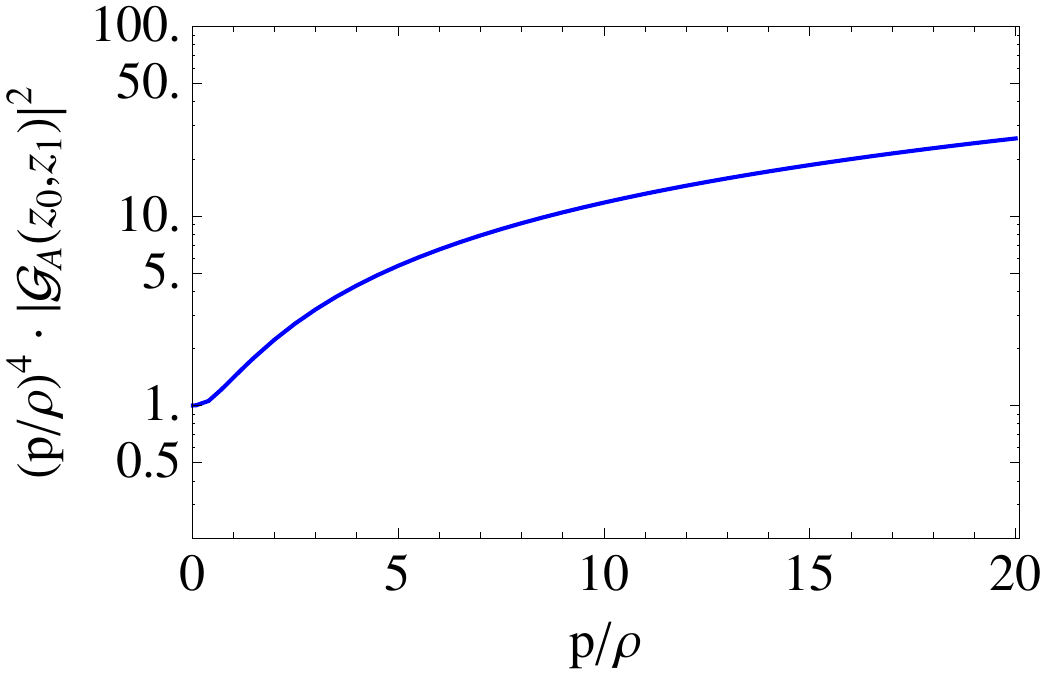}
\caption{\it Left (right) panel is $\sigma(q\bar q\to g^*\to Q\bar Q)/\sigma_{SM}(q\bar q\to g^0\to Q\bar Q)$, versus $p/\rho$, with $p\equiv\sqrt{\hat s}$ being the partonic energy, while $Q$ is a light (heavy) quark living on the UV (IR) brane. We have used $A_1 = 35$ and $c=1$.}
\label{fig:cross_section}
\end{figure} 
or a light quark living on the UV brane (as e.g.~$b_{L,R}$). In the right panel of Fig.~\ref{fig:cross_section} we show the case where $Q=t_R$, where we assume that $t_R$ is living on the IR brane. As fermions localized on the IR brane are strongly coupled to the KK modes, the relative cross-section, with respect to the SM one, increases with the partonic energy and can yield a sizeable departure from the SM prediction for large values of the partonic energy. We see from the right panel of Fig.~\ref{fig:cross_section} that the enhancement can be $\mathcal O(10)$ for  $\sqrt{\hat s}\simeq\mathcal O(10)\rho$. In the left panel of  Fig.~\ref{fig:cross_section} we show the case where $Q$ is a light quark localized toward the UV brane. We see that the enhancement with respect to the SM prediction is $\mathcal O(1)$ for any partonic energy and thus much more difficult to detect experimentally. Notice also that in the limit $p\to 0$ the processes are dominated by the gluon zero mode (an isolated pole on top of the gluon continuum) which makes, in this limit, $\sigma(q\bar q\to g^*\to Q\bar Q)/\sigma_{SM}(q\bar q\to g^0\to Q\bar Q)\to 1$, as can be seen from Figs.~\ref{fig:cross_section}.
\begin{figure}[htb]
\centering
\includegraphics[width=7.5cm]{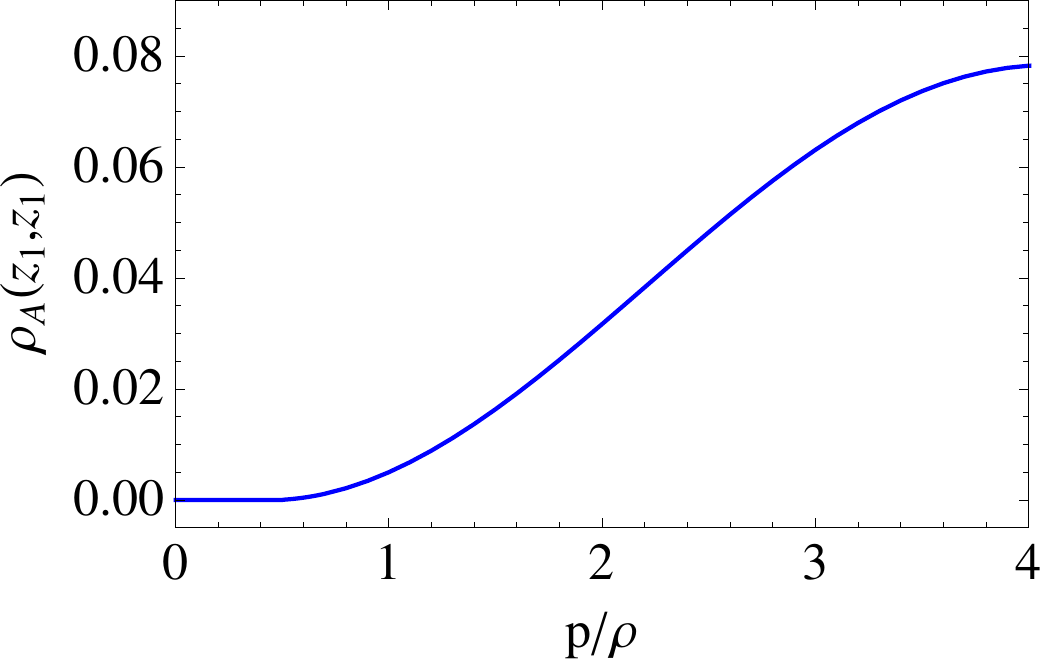} \hspace{0.15cm}
\includegraphics[width=7.5cm]{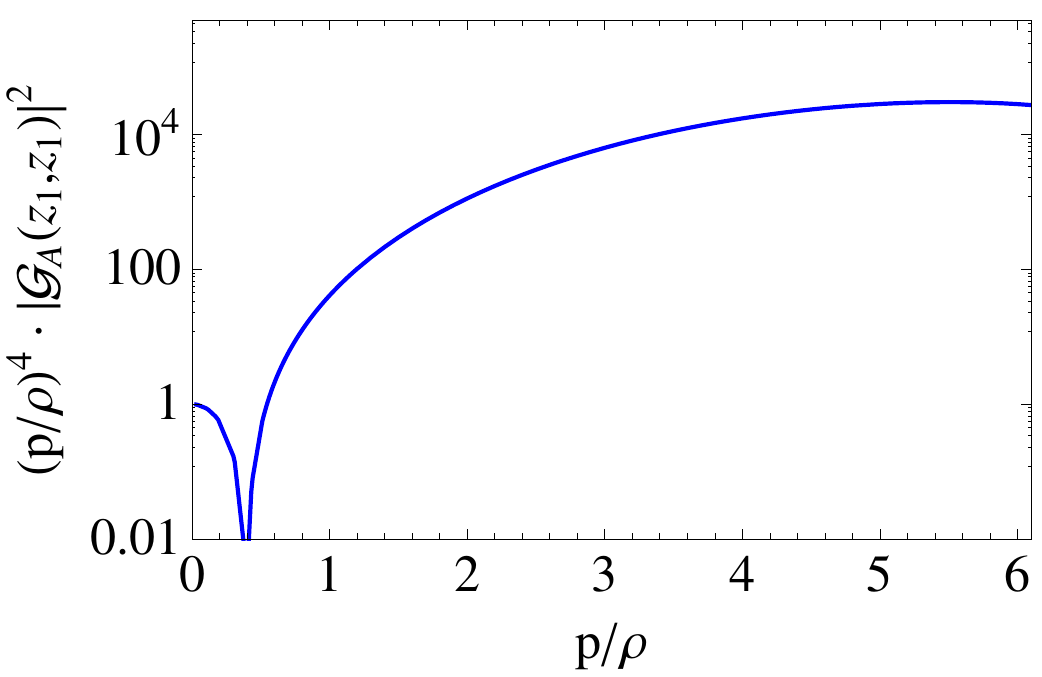} 
\caption{\it  Scale invariant spectral density (left panel) and squared absolute value of the Green's function (center panel) for gauge bosons, for the case $z=z_1$ and $z'=z_1$. We have used $A_1 = 35$ and $c=1$.
}
\label{fig:spectral_gaugebosony1y1}
\end{figure} 

Finally we will consider here the case of IR-brane-to-IR-brane spectral and Green's functions $\sigma_A(z_1,z_1;p)$ and $G_A(z_1,z_1;p)$. This case is relevant in processes where both the initial and final fermions are localized on the IR brane. This is the case for instance in models explaining the $R_{D^{(\ast)}}$ anomalies, where right-handed third generation fermions are localized on the IR brane~\cite{Carena:2018cow}. The relevant Green's function can contribute significantly to the process $\sigma(b_R \bar b_R\to g^*\to t_R\bar t_R)$, which is parton distribution function (PDF) suppressed, with respect to $\sigma(q \bar q\to g^*\to t_R\bar t_R)$, by the small amount of bottoms inside the proton, but enhanced, with respect to the latter, by the large coupling of the bottom to the KK modes, while the SM contribution $\sigma_{SM}(b_R\bar b_R\to g^{(0)}\to t_R \bar t_R)$ is suppressed. The scale invariant spectral function 
\be
\rho_A(z_1,z_1;p)\equiv (\rho^2/k) \sigma_A(z_1,z_1;p)
\ee
is shown in the left panel of Fig.~\ref{fig:spectral_gaugebosony1y1}, while the scale invariant Green's function, \\ $(\hat s/\rho^2)\mathcal G_A(z_1,z_1;p)$, where
\be
\mathcal G_A(z_1,z_1;p)\equiv \mathcal W(k/\rho) (\rho^2/k)G_A(z_1,z_1;p)
\ee
which measures the ratio $\sigma(b_R \bar b_R\to g^*\to t_R\bar t_R)/\sigma_{SM}(b_R\bar b_R\to g^{(0)}\to t_R \bar t_R)$ is shown in the right panel of Fig.~\ref{fig:spectral_gaugebosony1y1}. As we can see, the enhancement of the production through the gluon continuum can easily be $\mathcal O(100-1000)$, and so for large collider energies it can be significant and lead, in spite of the strong PDF suppression, to a strong deviation with respect to the SM predictions. Notice also that $\sigma(b_R \bar b_R\to g^*\to t_R\bar t_R)/\sigma_{SM}(b_R\bar b_R \to g^{(0)}\to t_R \bar t_R)\to 1$ in the limit $p\to 0$, as the process is dominated by the isolated zero mode of the gluon.

\section{Conclusions}
\label{sec:conclusions}

In this paper we have proposed a model with a warped extra dimension solving the hierarchy problem, and where the KK spectra of all particles (gauge bosons, fermions, graviton, radion, Higgs boson) are continua of states with a mass gap. In this sense it provides a modelization of the theory of unparticles in the presence of a mass gap at the TeV scale. The existence of such continua should modify the present searches of new physics, which are mainly concentrated on the presence of bumps in the invariant mass of final states, corresponding to isolated (and narrow) resonances. We have shown that the model predicts the existence of a metric singularity at a finite distance in proper coordinates $y=y_s$ (which corresponds in conformal coordinates to $z_s\to\infty$). The singularity is admissible as it supports finite temperature in the form of a black hole horizon. The space has as boundaries a UV brane and the singularity.  Moreover, we have introduced an IR brane to successfully trigger electroweak breaking, and which requires special jump conditions. The distance between branes and the singularity are fixed by the dynamics of a bulk Goldberger-Wise field with brane potentials breaking the conformal invariance. 

The gravitational background is AdS$_5$ near the UV brane (which allows to solve the hierarchy problem \textit{\` a la} Randall-Sundrum) and has strong breaking of conformality near the IR. In this way it behaves like the linear dilaton theory near the IR while it departs from it near the UV, where it behaves as the RS theory. This requirement is proved to be essential to solve the hierarchy problem in the conventional fashion, where the Planck scale is fundamental and the TeV scale is derived from it by the metric warp factor. Here we depart from  the 5D version of clockwork theories, based on Little String Theories, where the TeV scale is fundamental and the Planck scale is a derived one. As a consequence of the existence of two very different regimes, in the UV and IR regions, the equations of motion and Green's functions, which are computed in this paper, cannot be solved analytically, as was done in the model of Ref.~\cite{Csaki:2018kxb}, but all calculations have to be done numerically. In particular we have computed UV-brane-to-UV-brane Green's and spectral functions using holographic methods, and arbitrary Green's function by solving the non-homogeneous equations of motion. 

As a particular phenomenological application we have computed the brane-to-brane Green's function and seen how it can modify, at the LHC, the SM cross-section where a gluon is produced by Drell-Yan process and decays into a pair of heavy fermions $Q$ localized on the IR brane. The prediction is a smooth increase of the cross-section which depends on the partonic energy, and should produce an increase in the cross section $\sigma(pp\to Q\bar Q)$. Other, more detailed, phenomenological applications should be inspired to a large extent on unparticle phenomenology and we will postpone their study for a future work.

\vspace{0.5cm}
\section*{Acknowledgments}
%%%%%%%%%
We would like to thank O. Pujol\`as and L.L. Salcedo for
discussions. Special thanks are due to M. P\'erez-Victoria for very
interesting comments about the contents of the paper and the previous
literature on the subject. The work of EM is supported by the Spanish
MINEICO under Grant FIS2017-85053-C2-1-P, by the Junta de
Andaluc\'{\i}a under Grant FQM-225, by the Consejer\'{\i}a de
Conocimiento, Investigaci\'on y Universidad of the Junta de
Andaluc\'{\i}a and European Regional Development Fund (ERDF) under
Grant SOMM17/6105/UGR, and by the Spanish Consolider Ingenio 2010
Programme CPAN under Grant CSD2007-00042. The research of EM is also
supported by the Ram\'on y Cajal Program of the Spanish MINEICO under
Grant RYC-2016-20678. The work of MQ is partly supported by Spanish
MINEICO (Grant FPA2017-88915-P), by the Catalan Government under Grant
2017SGR1069, and by Severo Ochoa Excellence Program of MINEICO (Grant
SEV-2016-0588).

\bibliographystyle{JHEP}
\bibliography{refs}

\end{document}